\documentclass[a4paper,12pt]{article}
\usepackage{amsfonts}
\usepackage{color}

\usepackage{amsmath}

\newtheorem{theorem}{Theorem}
\newtheorem{corollary}{Corollary}
\newtheorem{definition}{Definition}
\newtheorem{proposition}{Proposition}
\newtheorem{remark}{Remark}
\newtheorem{lemma}{Lemma}
\newenvironment{acknowledgement}[1][Acknowledgement]{\textbf{#1.}}{}
\newenvironment{proof}[1][Proof]{\textbf{#1.} }{\ \rule{0.5em}{0.5em}}

\def\complex{{\mathbb C}}
\def\real{{\mathbb R}}

\def\z{{\mathbb Z}}

\def\bb{\cal B}
\def\cc{\cal C}

\def\hh{\cal H}

\def\kk{\cal K}
\def\ll{\cal L}
\def\mm{\cal M}

\def\rr{\cal R}

\def\tor{\mathbb T}

\begin{document}

\author{C.~Carmeli\thanks{C.~Carmeli,
Dipartimento di Fisica, Universit\`a di Genova, and I.N.F.N.,
Sezione di Genova, Via Dodecaneso~33, 16146 Genova, Italy. e-mail:
carmeli@ge.infn.it}, G. Cassinelli\thanks{G.~Cassinelli,
Dipartimento di Fisica, Universit\`a di Genova, and I.N.F.N.,
Sezione di Genova, Via Dodecaneso~33, 16146 Genova, Italy. e-mail:
cassinelli@ge.infn.it}, A.
Toigo\thanks{A.~Toigo,
Dipartimento di Fisica, Universit\`a di Genova, and I.N.F.N.,
Sezione di Genova, Via Dodecaneso~33, 16146 Genova, Italy. e-mail:
toigo@ge.infn.it},V.S. Varadarajan \thanks{V.S.Varadarajan, Department of 
Mathematics, University of California at Los Angeles, Box 951555 Los 
Angeles, CA 90095-1555, USA.
e-mail: vsv@math.ucla.edu}
}
\title{Unitary representations of super Lie
groups and applications to the classification and multiplet
structure of super particles}
\date{\today}

\maketitle


\maketitle

\begin{abstract}
It is well known that the category of super Lie groups (SLG) is equivalent to the category of super Harish-Chandra pairs (SHCP). Using this equivalence, we define the category of unitary representations (UR's) of a super Lie group. We give an extension of the classical inducing construction and Mackey imprimitivity theorem to this setting.
We use our results to classify the irreducible unitary
representations of semidirect products of super translation groups
by classical Lie groups, in particular of the super Poincar\'e
groups in arbitrary dimension. Finally we compare
our results with those in the physical literature on the structure
and classification of super multiplets.
\end{abstract}

\section{Introduction}
\label{intro}
The classification of free relativistic super
particles in SUSY quantum mechanics is well
known (see for example \cite{SS},\cite{FSZ}). It is based on the
technique of little (super)groups which, in the
classical context, goes back to Wigner and Mackey.
However the treatments of this question
in the physics literature make the
implicit assumption that the technique of
little groups remains valid in the SUSY
set up without any changes; in particular no
attempt is  made in currently available treatments
to exhibit
the SUSY transformations
explicitly for the super particle. The aim of
this paper is to remedy this situation by
laying a precise
mathematical foundation for the theory of
unitary representations of super Lie groups,
and then to
apply it to the case of the super Poincar\'e
groups. In the process of doing this we
clarify and extend the
results in the physical literature to
Minkowskian spacetimes of arbitrary dimension
$D\ge 4$ and
$N$-extended supersymmetry for arbitrary $N\ge 1$.

\medskip

Super Lie groups differ from classical Lie
groups in a fundamental way: one should think
of them as group-valued functors rather than
groups. As long as only finite dimensional
representations are being considered, there is
no difficulty in adapting the functorial theory
to study representations. However, this marriage
of the functorial approach with
representation theory requires modifications
when one considers infinite dimensional
representations. Our entire approach is based on two
observations. The first is to view a super Lie group
as a Harish-Chandra pair, namely a pair $(G_0, \mathfrak g)$
where $G_0$ is a classical Lie group, $\mathfrak g$ is a
super Lie algebra which is a $G_0$-module,
${\rm Lie}(G_0)=\mathfrak g_0$, and the action of
${\mathfrak g}_0$ on $\mathfrak g$ is the differential of the action of $G_0$.
That this is a valid starting point is justified by
the result \cite{DM} that the category of super Lie groups
is equivalent to the category of Harish-Chandra pairs.
This point of view leads naturally to define a
unitary representation of a super Lie group $(G_0, \mathfrak g)$
as a classical unitary representation of $G_0$ together with
a compatible infinitesimal unitary action of $\mathfrak g$. This is in fact very close
to the approach of the physicists.
The source of our second observation is more technical
and is the fact, which is a consequence
of the commutation rules, that the operators corresponding
to the odd elements of $\mathfrak g$ are in general
{\it unbounded\/} and so care is needed to work with them.
Our second observation is in fact a basic
result of this paper, namely, that the commutation rules
and the symmetry requirements that are implicit
in a supersymmetric theory force the unbounded odd
operators to be well behaved and lead to an essentially
unique way to define a unitary representation of a
super Lie group. Of course this aspect was not treated
in the physical literature, not only because
representations of the big super Lie groups were not considered,
but even more, because only finite dimensional
super representations of the little groups were considered,
where unbounded phenomena obviously do not occur. Our
treatment has the additional feature that it is able to
handle the construction of super particles with
{\it infinite spin} also.

\medskip

In the first section of the paper we treat the foundations
of the theory of unitary representations of a super
Lie group based on these two observations.
The basic result is Proposition \ref{p1.3.3}
which asserts that the odd operators are {\it essentially
self adjoint\/} on their domain and that the representation
of the Harish-Chandra pair,
extended to the space of $C^\infty$ vectors of the
representation of $G_0$, is unique. This result is
essential to everything we do and shows that the
formal aspects of representations of super Lie groups
already control all their analytic aspects. The
second section discusses the imprimitivity theorem
in the super context. Here an important assumption is
made, namely that the super homogeneous space on which
we have the system of imprimitivity is a {\it purely even\/}
manifold,
or equivalently, the sub super Lie group $(H_0, \mathfrak h)$
defining
the system of imprimitivity has the same
odd dimension as the ambient super Lie group,
i.e., $\mathfrak h_1=\mathfrak g_1$.
This restriction, although severe, is entirely adequate
for treating super Poincar\'e groups and more generally
a {\it wide class of  super semi direct products.\/} The main result is
Theorem~\ref{t2.3.1} which asserts that
the inducing functor is an equivalence of categories from
the category of unitary representations of $(H_0, \mathfrak h)$
to the category of super systems of imprimitivity based
on $G_0/H_0$. These two sections complete the
foundational aspects of this paper.

\medskip

The third section is concerned with applications and
bringing our treatment as close as possible to the
ones in the literature. We
consider a super semi direct product of a
super translation group with a classical Lie group $L_0$
acting on it. {\it No special assumption
is made about the action of
$L_0$ on $\mathfrak g_1$, so that the class of super Lie groups
considered is vastly larger than the ones treated in
the physical literature, where this action is always
assumed to be spinorial.\/} If $T_0$ is the
vector group which is the even part of the super
translation group, the theory developed in \S\S\ref{sec:2},\ref{sec:3} leads to
the result that the irreducible representations of
$(G_0, \mathfrak g)$ are in one-one correspondence
with {\it certain\/} $L_0$-orbits in the dual $T_0^\ast$ of $T_0$
together with {\it certain\/} irreducible representations of the
super Lie groups (little groups) which are
stabilizers of the points in the orbits. This is
the super version of the classical little group method
(Theorem \ref{t3.1.1}).

\medskip

It is from this point on that the SUSY theory acquires its
own distinctive flavor. In the first place, unlike the
classical situation, Theorem \ref{t3.1.1} stipulates that
not all orbits are allowed, only those belonging to a
suitable subset $T_0^+$. We shall call these orbits
{\it admissible.\/} These are the orbits
where the little super group admits an irreducible
unitary representation  which restricts to a character of $T_0$. These representations will be called {\it admissible}. These orbits satisfy a positivity
condition which we interpret as the condition of
positivity of energy. This condition is therefore necessary
for admissibility. However it requires some effort
to show that it is also sufficient for admissibility,
and then to determine all the irreducible
unitary representations of the little
super Lie group at $\lambda$ (Theorem \ref{t3.2.7}).

\medskip

The road to Theorem \ref{t3.2.7} is somewhat complicated.
Let $\lambda \in T_0^\ast$ be fixed. The classical stabilizer
of $\lambda$ is $T_0L^\lambda_0$, where $L^\lambda_0$
is the classical little group at $\lambda$,
namely the stabilizer of $\lambda$ in $L_0$.
The super stabilizer of $\lambda$ is the super Lie group
$S^\lambda$ defined by
$$
S^\lambda=(T_0L^\lambda_0, \mathfrak g^\lambda ),\qquad
\mathfrak g^\lambda =\mathfrak t_0\oplus
\mathfrak l^\lambda_0\oplus \mathfrak g_1.
$$
(By convention, the Lie algebra of a Lie group
is denoted by the corresponding gothic letter.)
Given $\lambda$,
there is an
associated  $L_0^\lambda$-invariant  quadratic form $\Phi_\lambda$ on $\mathfrak g_1$
which will be  nonnegative definite; the nonnegativity of
$\Phi_\lambda$ is the positive energy
condition mentioned earlier. This form need not be
strictly positive, but one can pass to the quotient
$\mathfrak g_{1\lambda}$ of
$\mathfrak g_1$ by its radical, and obtain a
positive definite quadratic vector space on which the classical
part $L^\lambda _0$ of the little group operates.
So we obtain a map
$$
j_\lambda : L^\lambda _0\longrightarrow
{\rm O}(\mathfrak g_{1\lambda }).
$$
Now $L^\lambda_0$ need not
be connected and so $j_\lambda$
need not map into ${\rm SO}(\mathfrak g_{1\lambda })$. Also,
since we are not making any assumption about the action of
$L_0$ on $\mathfrak g_1$, it is quite possible that
$\dim (\mathfrak g_{1\lambda})$ could be odd. We introduce
the subgroup
$$
L^\lambda_{00}=
\left\{
\begin{array}{ll}
j_\lambda ^{-1}({\rm SO}
(\mathfrak g_{1\lambda })) & \mbox{if } j_\lambda (L^\lambda _0)\not\subset {\rm SO}
(\mathfrak g_{1\lambda }) \mbox{ and } \dim (\mathfrak g_{1\lambda })\mbox{ is even}\\
L^\lambda _0 & \mbox{otherwise.}
\end{array}
\right.
$$
Then $L^\lambda_{00}$ is either the whole of $L^\lambda _0$ or a
(normal) subgroup of index $2$ in it.
Theorem \ref{t3.2.7} asserts that there is a functorial map
$$
r\longmapsto \theta _{r\lambda}
$$
which is an equivalence of categories from the category
of unitary projective representations $r$
of $L^\lambda_{00}$ corresponding to a certain canonical
multiplier $\mu_\lambda$, to the category of
admissible unitary representations of the super group $S^\lambda$.
In particular, the admissible irreducible unitary representations
of $S^\lambda$ correspond one-one to irreducible
unitary $\mu_\lambda$-representations of
$L^\lambda_{00}$. We shall now
explain how this correspondence is set up.

\medskip

The odd operators
of the super representation of the little group are
obtained from a self adjoint representation of the
Clifford algebra of $\mathfrak g_{1\lambda}$. Since any
such is a multiple of an essentially unique one, we begin
with an {\it irreducible\/} self adjoint representation. Let us
call it $\tau_\lambda$. In the space of $\tau_\lambda$
one can define in an essentially unique manner
a projective representation
$\kappa_\lambda$ of $L^\lambda_0$ that intertwines
$\tau_\lambda $
and its transforms by elements of $L^\lambda _0$:
$$
\kappa_\lambda (t)\tau_\lambda(X)\kappa_\lambda (t)^{-1}
=\tau_\lambda (tX)\eqno (\ast ).
$$
The class of the multiplier $\mu_\lambda$
in $H^2(L^\lambda _0, \tor)$ is
then uniquely determined. We can normalize
$\mu_\lambda$ so that it takes only $\pm 1$-values. Starting from any
unitary $\mu_\lambda$-representation $r$ of $L^\lambda_{00}$
we build the $\mu_\lambda$-representation
$$
{\rm Ind}(r)
$$
of $L^\lambda_0$ induced by $r$. Then the
admissible unitary representation $\theta_{r\lambda}$
of $S^\lambda$ corresponding
to $r$ is given by
$$
\theta _{r\lambda}=\left(e^{i\lambda (t)}
{\rm Ind}(r)\otimes \kappa_\lambda ,
1\otimes\tau_\lambda\right);
$$
the fact that $\mu_\lambda$ is
$\pm 1$-valued implies that $\theta_{r\lambda}$
is an ordinary
rather than a projective representation.
The representation $\Theta _{r\lambda }$ of the super
Lie group $(G_0, \mathfrak g)$ induced by $\theta_{r\lambda }$
from $S^\lambda$
is then irreducible if $r$ is irreducible. Theorem \ref{t3.2.8}
asserts that all the
irreducible unitary representations of the super Lie group
are parametrized bijectively as above by
the irreducible representations of $L^\lambda _{00}$,
thus giving the
super version of the classical theory.

\medskip

The representation $\kappa_\lambda$ is therefore at the heart
of the theory of irreducible unitary representations of
super semi direct products. It is finite dimensional;
in fact it is the lift via $j_\lambda$ to $L^\lambda _0$
of the spin
representation of the quadratic vector space. It is
dependent only on $\lambda$. Clearly, the
representation ${\rm Ind\/}(r)\otimes \kappa_\lambda$
of $L^\lambda_0$ will not in general
be irreducible even if $r$ is. The irreducible constituents
of ${\rm Ind\/}(r)\otimes \kappa_\lambda$ then define, via
the classical procedure of inducing,
irreducible unitary representations of $G_0$ which are the
constituents of the even part of the
full super representation. This
family of irreducible representations of $G_0$ is the
{\it multiplet\/} of the irreducible representation
of the super semi direct product in question.
If $\mu_\lambda$ is trivial, we can choose $r$ to be trivial, and the corresponding 
multiplet
 is the {\it fundamental multiplet.\/} Thus
$\kappa_\lambda$ determines the entire correspondence in
a simple manner.

\medskip

In \S\ref{ssec:3.3} we discuss the case of the super Poincar\'e groups.
We consider spacetimes of Minkowski signature and of
arbitrary dimension $D\ge 4$ together with
$N$-extended supersymmetry for arbitrary $N\ge 1$.
In this case, the groups
$L^\lambda_0$ are all connected and $L^\lambda_0=
L^\lambda_{00}$. Moreover, the multiplier $\mu_\lambda$
becomes trivial, so that {\it $\kappa_\lambda$
becomes an ordinary representation\/}
(Lemma \ref{l3.2.4}). Hence
$$
\theta_{r\lambda}=\left( e^{i\lambda (t)}r\otimes
\kappa_\lambda,1\otimes \tau_\lambda\right).
$$
Thus in
this case we finally reach the conclusion that
the super particles are parametrized by the admissible
orbits and irreducible unitary representations
of the stabilizers
of the classical little groups, exactly as in
the classical theory. The
positive energy condition $\Phi_\lambda \ge 0$
becomes just that $\lambda $,
which we replace by $p$ to display the fact that it is a momentum
vector, lies in the closure of the forward light cone. Thus
the orbits of imaginary mass are excluded by
supersymmetry (Theorem \ref{t3.3.5}). Our approach enables us to
handle super particles with {\it infinite spin\/}
in the same manner as
those with finite spin because of the result that the odd
operators of the little group are {\it bounded\/} (Lemma \ref{l3.2.1}).

\medskip

Finally, in \S\ref{ssec:3.4} we give the explicit determination of
$\kappa_p$ when the spacetime has arbitrary dimension $D\ge 4$
and we have $N$-extended supersymmetry.
The results in the physical literature are in general only
for $D=4$.

 \medskip

The literature on supersymmetric representation
theory is very
extensive. We have been particularly influenced by \cite{SS},
\cite{F1},\cite{F2},\cite{FSZ}. 
In  \cite{Do85}, \cite{Do86}, \cite{Do} Dobrev et al. discuss representations of superconformal groups induced from a maximal parabolic subgroup.
The papers
\cite{SS} gave the structure of $\kappa_p$  for $N=1$ supersymmetry,
while \cite{FSZ} gave
a very complete treatment of the structure of $\kappa_p$
in the case of extended supersymmetry when $D=4$, including
the case of central charges.

\bigskip\noindent

\section{Super Lie groups and their
unitary representations}
\label{sec:1}
\subsection{Super Hilbert spaces}
\label{ssec:1.1}
 All sesquilinear forms are
linear in the first argument
and conjugate linear in the
second. We use the usual terminology of super geometry
as in \cite{DM},\cite{V}.  A {\em super Hilbert space}
(SHS) is a super vector space
${\hh}={\hh}_0\oplus {\hh}_1$ over ${\complex}$ with a
scalar product $({\cdot}\ ,\ {\cdot})$ such that
${\hh}$ is a Hilbert space
under $({\cdot}\ ,\ {\cdot})$, and ${\hh}_i (i=0,1)$
are mutually orthogonal closed linear subspaces.
If we define
$$
\langle x, y\rangle=
\begin{cases}
0 & \mbox{ if } x \mbox{ and } y \mbox{ are of opposite parity}\cr
(x, y) & \mbox{ if } x \mbox{ and } y \mbox{ are even } \cr
i(x,y) & \mbox{ if } x \mbox{ and } y \mbox{ are odd }  \cr
\end{cases}
$$
then $\langle x,y\rangle$ is an even super Hermitian form with
$$
\langle y,x\rangle=
(-1)^{p(x)p(y)}\overline {\langle x,y\rangle},\
\langle x,x\rangle >0 (x\not=0\ {\rm  even }),\
i^{-1}\langle x,x\rangle >0
(x\not=0\  {\rm  odd }).
$$
If $T({\hh}\rightarrow {\hh})$ is a bounded linear operator,
we denote by $T^\ast$ its Hilbert
space adjoint
and by $T^{\dag} $ its super adjoint given by $\langle Tx,y\rangle =
(-1)^{p(T)p(x)}\langle x, T^{\dag} y\rangle $.
Clearly $T^{\dag}$ is bounded, $p(T)=p(T^{\dag})$, and
$T^{\dag} =T^{\ast}$ or $-iT^{\ast}$ according as $T$
is even or odd. For unbounded $T$ we define
$T^{\dag}$ in terms of $T^{\ast}$ by the above formula.
These definitions are equally consistent if
we use $-i$ in place of $i$.
But our convention is as above.\\
\subsection{SUSY quantum mechanics}
\label{ssec:1.2}
In SUSY
quantum mechanics in a SHS ${\hh}$, it is usual to stipulate that the
Hamiltonian $H=X^2$ where
$X$ is an odd operator \cite{W}; it is customary to argue that this
implies that $H\ge 0$ (positivity
of energy); but this is true only if we know that $H$ is essentially self
adjoint on the domain of $X^2$. We shall now prove two lemmas which
clarify this situation and will play a crucial role when we consider
systems
with a super Lie group of symmetries.

\medskip

If $A$ is a linear operator on $\mathcal{H}$, we denote by $D\left( A \right)$
its domain. We always assume that $D\left( A \right)$ is dense in $\mathcal{H}$, and 
then refer to it as {\em densely defined}.
 We write
$A\prec B$ if $D(A)\subset D(B)$ and $B$
restricts to $A$ on $D(A)$; $A$ is symmetric iff $A\prec A^\ast$, and
then $A$ has a closure $\overline A$.
$A\prec B\Longrightarrow B^\ast \prec A^\ast$. If $A$ is symmetric
we say that it is essentially
self adjoint if $\overline A$ is self adjoint; in this case $A^\ast
=\overline A$.
If $A$ is symmetric and $B$ is a symmetric  extension of $A$,
then $A\prec B\prec A^\ast$; in fact $A\prec A^\ast$ and $A\prec B\prec
B^\ast $,
and so $B^\ast \prec A^\ast$ and $A\prec B\prec B^\ast \prec A^\ast$. If
$A$ is
self adjoint and $L\subset D(A)$, we say that $L$ is a {\it core\/} for
$A$
if $A$ is the closure of its restriction to $L$. A vector $\psi \in {\hh}$
is analytic for a symmetric operator $H$ if $\psi \in D(H^n)$ for all $n$
and
the series $\sum _nt^n(n!)^{-1}||H^n\psi ||<\infty $ for some $t>0$.
It is a well known result
of Nelson \cite{N} that if $D\subset D(H)$ and contains a dense set of
analytic vectors, then
$H$ is essentially self adjoint on $D$. In this case $\psi \in D(H)$ is
analytic for $H$ if and only if $t\longmapsto e^{itH}\psi $ is analytic in
$t\in {\real}$.
If $A$ is self adjoint, then $A^2$, defined on the domain $D(A^2)=
\{\psi \ \big|\ \psi, A\psi \in D(A)\}$, is self adjoint;
this is well known and follows easily from the spectral theorem.

\begin{lemma}
\label{l1.2.1}
Let $H$ be a self adjoint operator on ${\hh}$
and
$U(t)=e^{itH}$ the
corresponding one parameter unitary group. Let ${\bb}\subset D(H)$ be a
dense $U$-invariant
linear
subspace. We then have the following.
\begin{itemize}

\smallskip \item[{\rm (}i{\rm )}] ${\bb}$ is a core for $H$.

\smallskip \item[{\rm (}ii{\rm )}] Let $X$ be a symmetric operator
with
${\bb}\subset D(X)$ such that
$X{\bb}\subset D(X)$ and $\left. X^2 \right|_{\bb} = \left. H \right|_{\bb} $. Then
$\left. X \right|_{\bb}$ is essentially self
adjoint, $\overline{\left. X \right|_{\bb}} = \overline{X} $
and $\overline X^2=H$.
\end{itemize}
In particular, $H\ge 0$,
$D(H)\subset D(\overline X)$. Finally,
 these results are valid if we only assume that ${\bb}$ is invariant under
$H$ and
contains a dense set of analytic vectors.
\end{lemma}
\begin{proof} Let $H_1=H\big |_{\bb}$. We must show that if
$L(\lambda ) (\lambda \in {\complex})$ is the subspace
of $\psi $ such that $H_1^\ast \psi =\lambda \psi $, then $L(\lambda )=0$
if
$\Im (\lambda )\not=0$.
Now $L(\lambda )$ is a closed subspace. Moreover, as $H_1$ is invariant
under $U$,
so is $H_1^\ast $
and so $L(\lambda )$ is invariant under $U$ also. So the vectors in
$L(\lambda)$ that are
$C^\infty $ for $U$ are dense in $L(\lambda )$ and so it is enough to
prove
that $0$ is the only
$C^\infty $ vector in $L(\lambda )$. But $H=H^\ast \prec H_1^\ast$ while
the $C^\infty $ vectors
for $U$ are all in $D(H)$, and so if $\psi $ is a $C^\infty $
vector in $L(\lambda )$, $H\psi =H_1^\ast \psi =\lambda \psi$.
This is a contradiction since $H$ is
self adjoint
and so all its eigenvalues are real. This proves (i).

\medskip
Let $X_1 = X\big |_{\bb}$. Clearly, $X_1$ is symmetric on ${\bb}$. It is enough to show that
$X_1$ is essentially self adjoint and $\overline{X_1}^2=H$, since in this case
$X_1 \prec X \prec X_1^\ast = \overline{X_1}$ and hence $\overline{X}=
\overline{X_1}$. We have $X_1^2 =H_1$. So $H\ge
0$ on ${\bb}$
and hence
$H\ge 0$ by (i). Again it is a question of showing that for $\lambda \in
{\complex}$
with $\Im (\lambda )\not=0$, we must have $M(\lambda )=0$ where $M(\lambda
)$ is
the eigenspace for
$X_1^\ast $ for the eigenvalue $\lambda $. Let $\psi \in M(\lambda )$.
Now, for $\varphi \in {\bb}$,
$$
(X_1^2\varphi ,\psi )=(X_1 \varphi ,X_1^\ast \psi )=\overline \lambda (X_1\varphi
,\psi )=
\overline \lambda ^2(\varphi ,\psi )=(\varphi ,\lambda ^2\psi ).
$$
Hence $\psi \in D((X_1^2)^\ast )$ and $(X_1^2)^\ast \psi =
\lambda ^2\psi$. But $X_1^2=H_1$ and so $(X_1^2)^\ast =H_1^\ast =H$ by (i). So
$H\psi =\lambda ^2\psi$. Hence $\lambda ^2$ is real and $\ge 0$. This
contradicts
the fact that $\Im (\lambda )\not=0$. Furthermore,
$X_1^2\prec \overline{X_1}^2$ and so $\overline{X_1}^2=(\overline{X_1}^2)^\ast
\prec (X_1^2)^\ast=H_1^\ast =H$. On the other hand, as $\overline{X_1}^2$ is
closed,
$H=\overline H_1=\overline{X_1^2}\prec \overline{X_1}^2$. So $H=\overline{X_1}^2
= \overline
X^2$. This
means that $D(H)\subset D(\overline X)$.

\medskip
Finally, let us assume that
$H{\bb}\subset {\bb}$ and that ${\bb}$ contains
a dense set
of analytic vectors for $H$. Clearly ${\bb}$ is a core for $H$.
If $\psi \in {\bb}$ is analytic for $H$ we have $X^{2n}\psi = H^n \psi \in {\bb}$
and $X^{2n+1}\psi \in D(X)$ by assumption, and
$$
||X^n\psi ||^2=|(H^n\psi ,\psi )|\le ||\psi ||||H^n\psi ||\le M^n n!
$$
for some $M>0$ and all $n$. Thus $\psi $ is analytic for $X$ and so its
essential
self adjointness is a consequence of the theorem of Nelson.
The rest of the argument is unchanged.
\end{proof}
\begin{lemma}
\label{l1.2.2}
 Let $A$ be a self adjoint operator in ${\hh}$.
Let $M$ be a
smooth {\rm (}resp. analytic{\rm )} manifold
and $f(M\longrightarrow {\hh})$ a smooth {\rm (}resp. analytic{\rm )} map.
We
assume that {\rm (}i{\rm )}
$f(M)\subset
D(A^2)$ and {\rm (}ii{\rm )} $A^2f: m\longmapsto A^2f(m)$
is a smooth {\rm (}resp. analytic{\rm )} map of $M$
into ${\hh}$. Then $Af: m\longmapsto Af(m)$
is a smooth {\rm (}resp. analytic{\rm )} map of $M$ into ${\hh}$.
Moreover, if $E$ is any smooth differential operator on $M$, $(Ef)(m)\in
D(A^2)$
for all $m\in M$, and
$$
E(A^2f)=A^2Ef,\qquad E(Af)=AEf.
$$
\end{lemma}
\begin{proof} It is standard that if $g(M\longrightarrow {\hh})$
is smooth (resp. analytic)
and $L$ is a bounded linear operator on ${\hh}$, then $Lg$ is a
smooth (resp. analytic) map. We have
$$
A\psi =A(I+A^2)^{-1}(I+A^2)\psi ,\qquad \psi \in D(A^2).
$$
Moreover $A(I+A^2)^{-1}$ is a bounded operator. Now $(I+A^2)f$ is
smooth (resp. analytic) and so it is immediate from the above that $Af$ is
smooth (resp. analytic).

\medskip
For the last part we assume that $M$ is an open set in ${\real}^n$ since
the result is clearly local. Let $x^i (1\le i\le n)$ be the coordinates
and let
$\partial _j=\partial /\partial x^j, \partial ^\alpha =\partial _1^{\alpha
_1}\dots
\partial _n^{\alpha _n}$ where $\alpha =(\alpha
_1, \dots , \alpha _n)$. It is enough to prove that
$$
(\partial ^\alpha f)(m)\in D(A^2),\qquad
A^2\partial ^\alpha f=\partial ^\alpha (A^2f), \qquad
A\partial ^\alpha f=\partial ^\alpha (Af).
$$

\medskip
We begin with a simple observation. Since $(A^2\psi ,\psi )=||A\psi ||^2$
for all
$\psi \in D(A^2)$, it follows that whenever $\psi _n\in D(A^2)$ and
$(\psi _n)$ and
$(A^2\psi _n)$ are
Cauchy sequences, then $(A\psi _n)$ is also a Cauchy sequence;
moreover, if $\psi =\lim _n\psi _n$, then $\psi \in D(A^2)$
and $A^2\psi =\lim _nA^2\psi _n$, $A\psi =\lim _nA\psi _n$.
This said, we shall prove the above formulae by induction on
$|\alpha |=\alpha _1+\dots +\alpha _n$. Assume them for $|\alpha |\le r$
and fix
$j, 1\le j\le n$. Let $g=\partial ^\alpha f, |\alpha |=r$. Let
$$
g_h(x)={1\over h}\left (g(x^1,\dots ,x^j+h,\dots , x^n)-g(x^1,\dots
,x^n)\right )
\qquad (h \hbox { is in } j^{\rm th} \hbox { place}).
$$
Then, as $h\to 0$, $g_h(x)\to \partial _j\partial ^\alpha f(x)$
while $A^2g_h(x)=(\partial ^\alpha A^2f)_h(x)\to
\partial _j\partial ^\alpha A^2f(x)$, and
$Ag_h(x)=(\partial ^\alpha Af)_h(x)\to\partial _j\partial ^\alpha Af(x)$.
From the observation made above
we have $\partial _j\partial ^\alpha f(x)\in D(A^2)$ and $A^2$ and $A$ map
it respectively into $\partial _j\partial ^\alpha A^2f(x)$ and
$\partial _j\partial ^\alpha Af(x)$.
\end{proof}
\begin{definition} For self adjoint operators $L, X$ with $L$
bounded, we write $L\leftrightarrow X$ to mean that $L$
commutes with the spectral projections of $X$.
\end{definition}
\begin{lemma}
\label{l1.2.3}
 Let $X$ be a self adjoint operator and
${\bb}$ a dense subspace of ${\hh}$ which is a core for $X$ such
that $X{\bb}\subset {\bb}$. If $L$ is a
bounded self adjoint operator such that $L{\bb}\subset {\bb}$, then
the following are equivalent: {\rm (}i{\rm )} $LX=XL$ on ${\bb}$
{\rm (}ii{\rm )} $L\overline X=\overline XL$
on $D(\overline X)$ (this carries with it the inclusion
$LD(\overline X)\subset
D(\overline X)$)
{\rm (}iii{\rm )} $L\leftrightarrow \overline X$.
In this case $e^{itL}\overline X
=\overline Xe^{itL}$ for all $t\in {\real}$.
\end{lemma}
\begin{proof}
(i) $\Longleftrightarrow $ (ii). Let $b\in
D(\overline X)$.
Then there is a sequence $(b_n)$ in ${\bb}$ such that
$b_n\to b, Xb_n\to \overline Xb$.
Then $XLb_n=LXb_n\to L\overline Xb$. Since $Lb_n\to Lb$ we infer that
$Lb\in D(\overline X)$ and $\overline XLb=L\overline Xb$. This proves
(i) $\Longrightarrow $ (ii). The reverse implication is trivial.

\medskip\noindent
(ii) $\Longrightarrow $ (iii). We have $L^n\overline
Xb=\overline XL^nb$ for all
$b\in D(\overline X), n\ge 1$. So $e^{itL}\overline Xb=
\sum _n((it)^n/n!)\overline XL^nb$.
If $s_N=\sum _{n\le N}((it)^n/n!)L^nb$, then
$s_N\to e^{itL}b, \overline Xs_N
\to e^{itL}\overline Xb$. So $e^{itL}b\in D(\overline X)$
and $\overline Xe^{itL}b=
e^{itL}\overline Xb$. If $U(t)=e^{itL}$, this means
that $U(t)\overline XU(t)^{-1}
=\overline X$ and so, by the uniqueness
of the spectral resolution of $X$, $U(t)$ commutes with the spectral
projections of $\overline X$.
But then $L\leftrightarrow \overline X$.

\medskip\noindent
(iii) $\Longrightarrow $ (i). Under (iii) we have
$U(t)\overline XU(t)^{-1}=
\overline X$ or $\overline XU(t)b=U(t)\overline Xb$
for $b\in D(\overline X)$, $U(t)b$ being in $D(\overline X)$ for all
$t$. Let $b_t=(it)^{-1}(U(t)b-b)$. Then $\overline Xb_t=(it)^{-1}
(U(t)\overline Xb-\overline Xb)$. Hence, as $t\to 0$, $b_t\to Lb$
while $\overline Xb_t\to L\overline Xb$. Hence $Lb\in D(\overline X)$
and $\overline XLb=L\overline Xb$. Thus we have (ii), hence (i).
\end{proof}
\subsection{Unitary representations of super Lie groups}
\label{ssec:1.3}
We take the point of view \cite{DM} that a
super Lie group (SLG) is a {\em super Harish-Chandra pair} $(G_0, \mathfrak g)$ that is 
 a pair $(G_0, \mathfrak g)$
where $G_0$ is a classical Lie group, $\mathfrak g$ is a
super Lie algebra which is a $G_0$-module,
${\rm Lie}(G_0)=\mathfrak g_0$, and the action of
${\mathfrak g}_0$ on $\mathfrak g$ is the differential of the action of $G_0$.
The notion of morphisms between
two super Lie groups in the above sense is the obvious one from which
it is easy to see what is meant by a finite dimensional
representation of a  SLG $(G_0,
\mathfrak g)$:  it is a triple $(\pi _0, \pi , V)$ where $\pi _0$ is an
even representation of $G_0$
in a super vector space $V$ of finite dimension over ${\complex}$, i.e., a
representation such that $\pi _0(g)$ is even for all $g\in G_0$; $\pi $
is a representation of
the super Lie algebra $\mathfrak g$ in $V$ such that $\pi \big|_{\mathfrak
g_0}=d\pi _0$; and
$$
\pi (gX)=\pi _0(g)\pi (X)\pi _0(g)^{-1},\qquad g\in G_0, X\in \mathfrak g_1.
$$
If $V$ is a SHS and $\pi (X)^{\dag} =-\pi (X)$ for all $X\in \mathfrak g$, we
say that
$(\pi _0, \pi ,V)$ is a unitary representation (UR) of the SLG $(G_0,
\mathfrak g)$.
The condition on $\pi $
is equivalent to saying that $\pi_0$ is a unitary representation
of $G_0$ in the usual sense and $\pi (X)^{\ast}=-i\pi (X)$ for all $X\in
\mathfrak g_1$. It is
then clear that a finite dimensional UR of $(G_0, \mathfrak
g)$ is a triple $(\pi _0, \pi , V)$, where
\begin{itemize}

\smallskip\item[(a)] $\pi _0$ is an even unitary representation of $G_0$
is a SHS $V$;

\smallskip\item[(b)] $\pi $ is a linear map of $\mathfrak g_1$ into the space
$\mathfrak g\mathfrak l(V)_1$
of odd endomorphisms of $V$ with $\pi (X)^\ast =-i\pi (X)$ for all
$X\in \mathfrak g_1$;

\smallskip\item[(c)] $d\pi _0([X,Y])=\pi (X)\pi (Y)+\pi (Y)\pi (X)$ for
$X,Y\in \mathfrak g_1$;

\smallskip\item [(d)] $\pi (g_0X)=\pi _0(g_0)\pi (X)\pi _0(g_0)^{-1}$ for
$X\in \mathfrak g_1, g_0\in G_0$.
\end{itemize}
Let
$$
\zeta =e^{-i\pi /4}, \qquad \rho (X)=\zeta \pi (X).
$$
Then, we may replace $\pi (X)$ by $\rho (X)$ for $X\in \mathfrak g_1$; the
condition (b)
becomes the requirement that $\rho (X)$ is
self adjoint for all $X\in \mathfrak g_1$, while the commutation
rule in  condition (c)
becomes
$$
-id\pi _0([X,Y])=\rho (X)\rho (Y)+\rho (Y)\rho (X), \qquad
X,Y\in \mathfrak g_1.
$$

 \medskip
 If we want to extend this definition to the
infinite dimensional context it is necessary to
take into account the fact that the $\pi (X)$ for $X\in \mathfrak g_1$ will in
general
 be unbounded;
indeed, from (c) above we find that $d\pi_0([X,X])=2\pi (X)^2$, and as
$d\pi _0$ typically
takes elements of $\mathfrak g_0$ into
unbounded operators, the $\pi (X)$ cannot be bounded. So the concept of a
UR of a SLG in
the infinite dimensional case must be
formulated with greater care to take into account the domains of
definition of the
$\pi (X)$ for $X\in \mathfrak g_1$. In the physics literature this aspect is
generally
ignored. We shall prove below that
contrary to what one may
expect, the domain restrictions can be  formulated with great freedom, and
the formal
and rigorous pictures are essentially the
same.  In particular, the concept of a
UR of a super Lie group is a very stable one and allows great flexibility
of handling.

\medskip
If $V$ is
a super vector space (not necessarily finite dimensional), we write
${\bf End }(V)$ for the super
algebra of all endomorphisms of $V$. If $\pi _0$ is a unitary
representation of $G_0$
in a Hilbert space ${\hh}$, we write $C^\infty (\pi _0)$ for the subspace
of
differentiable vectors in ${\hh}$ for $\pi _0$. We denote by $C^\omega (\pi _0)$ the subspace of analytic vectors of $\pi _0$.
Here we recall that a vector $v\in\mathcal{H}$ is called a {\em differentiable vector} (resp. {\em analytic vector}) for $\pi_0$
 if the map
$g\mapsto \pi_0(g)v$ is smooth (resp. analytic). 
If ${\hh}$ is a SHS and
$\pi _0$ is even,
then $C^\infty (\pi _0)$ and $C^\omega (\pi _0)$
are $\pi _0$-invariant dense linear super subspaces. We also need
the following fact which is standard but we shall
give a partial proof because the argument will be used again later.
\begin{lemma}
\label{l1.3.1}
 $C^\infty (\pi _0)$ and $C^\omega (\pi _0)$ are stable under $d\pi
_0(\mathfrak g_0)$.
For any $Z\in \mathfrak g_0$, $id\pi _0(Z)$ is essentially
self adjoint both on $C^\infty (\pi _0)$ and on $C^\omega (\pi _0)$;
moreover, for any $Z_1, \dots ,Z_r\in \mathfrak g_0$ and $\psi \in C^\infty (\pi_0)$, {\rm (}resp.~$\psi \in C^\omega (\pi_0)${\rm )} the map
$g\longmapsto d\pi _0(Z_1)\dots d\pi _0(Z_r)\pi _0(g)\psi $ is $C^\infty
$ {\rm (}resp.~analytic{\rm )}.
\end{lemma}
\begin{proof}
 We prove only the second statement. That $id\pi _0(Z)$ is
essentially self adjoint on $C^\infty (\pi _0)$ and on $C^\omega (\pi _0)$ is immediate
from Lemma \ref{l1.2.1}. Using the adjoint representation
we have, for any $Z\in \mathfrak g_0$, $gZ=\sum _ic_i(g)W_i$ for $g\in G_0$
where the
$c_i$ are analytic functions on $G_0$ and $W_i\in \mathfrak g_0$. Hence,
as
$$d\pi_0 (Z)\pi_0(g)=\pi_0(g)d\pi_0(g^{-1}Z),
$$
we can write $d\pi _0(Z_1)\dots d\pi _0(Z_r)\pi _0(g)\psi $ as a linear
combination
with analytic coefficients of $\pi _0(g)d\pi _0(R_1)\dots d\pi _0(R_r)\psi $
for suitable $R_j\in \mathfrak g_0$. The result is then immediate.
\end{proof}
\begin{definition}
 A {\em unitary representation} (UR) of a SLG $(G_0, \mathfrak g)$ is a triple $(\pi _0,
\rho , {\hh}),
{\hh}$ a SHS,
with the following properties.
\begin{itemize}
\item[(a)] $\pi _0$ is an even UR of $G_0$ in ${\hh}$;
\item[(b)] $\rho (X\longmapsto \rho (X))$ is a linear map of
$\mathfrak g_1$ into
${\bf End}(C^\infty (\pi _0))_1$ such that
\begin{itemize}

\smallskip\item [(i)] $\rho (g_0X)=\pi _0(g_0)\rho (X)\pi _0(g_0)^{-1}
\ (X\in
\mathfrak g_1, g_0\in G_0)$,

\smallskip\item [(ii)] $\rho (X)$ with domain $C^\infty (\pi _0)$ is
symmetric
for all $X\in \mathfrak g_1$,

\smallskip\item [(iii)] $-id\pi _0([X,Y])=\rho (X)\rho (Y)+
\rho (Y)\rho (X) \ (X,Y\in \mathfrak g_1)$ on $C^\infty (\pi _0)$.
\end{itemize}
\end{itemize}
\end{definition}
\begin{proposition}
\label{p1.3.2}
 If $(\pi _0, \rho , {\hh})$ is a {\rm UR}
of the {\rm SLG} $(G_0, \mathfrak g)$, then $\rho (X)$ with domain
$C^\infty (\pi _0)$ is essentially self adjoint for all $X\in \mathfrak g_1$.
Moreover
$$
\pi : X_0+X_1\longmapsto d\pi _0(X_0)+\zeta ^{-1}\rho (X_1)\qquad (X_i\in
\mathfrak g_i)
$$
is a representation of $\mathfrak g$ in $C^\infty (\pi _0)$.
\end{proposition}
\begin{proof}
 Let $Z=(1/2)[X,X]$. We apply Lemma \ref{l1.2.1} with $U(t)=
\pi _0(\exp tZ)=e^{itH}, {\bb}=C^\infty (\pi _0)$. Then
$H=-id\pi _0(Z)=\rho (X)^2$ on $C^\infty (\pi _0)$. We conclude
that $H$ and
$\rho (X)$ are essentially self adjoint on $C^\infty (\pi _0)$
and that $H=\overline {\rho (X)}^2$; in particular, $H\ge 0$.
For the second assertion the
only non obvious statement
is that for
$Z\in \mathfrak g_0, X\in \mathfrak g_1, \psi \in C^\infty (\pi _0)$,
$$
\rho([Z,X])\psi =d\pi _0(Z)\rho (X)\psi -\rho (X)d\pi _0(Z)\psi .
$$
Let $g_t=\exp (tZ)$ and let $(X_k)$ be a basis for $\mathfrak g_1$. Then
$gX=\sum _kc_k(g)X_k$ where the $c_k$ are smooth functions on $G_0$. So
$$
\pi _0(g_t)\rho (X)\psi =\rho (g_tX)\pi _0(g_t)\psi =
\sum _kc_k(g_t)\rho (X_k)\pi _0(g_t)\psi.
$$
Now $g\longmapsto \pi _0(g)\psi $ is a smooth map into $C^\infty (\pi
_0)$.
On the other hand, if $H_k=-(1/2)[X_k,X_k]$, we have $id\pi
_0(H_k)=\rho(X_k)^2$
on $C^\infty (\pi _0)$,
so $\pi _0(g)\psi \in D(\rho (X_k)^2)$, and by Lemma \ref{l1.3.1},
$\rho (X_k)^2
\pi _0(g)\psi =id\pi _0(H_k)\pi _0(g)\psi $ is smooth in $g$.
Lemma \ref{l1.2.2} now applies
and shows that $\rho (X_k)\pi _0(g_t)\psi $ is smooth in $t$ and
$$
\left ({d\over dt}\right )_{t=0}\rho (X_k)\pi _0(g_t)\psi =\rho (X_k)d\pi
_0(Z)\psi.
$$
Hence
$$
d\pi _0(Z)\rho (X)\psi =\sum _kc_k(1)\rho (X_k)d\pi _0(Z)\psi +\sum
_k(Zc_k)(1)\rho (X_k)\psi .
$$
Since
$$
[Z,X]=\sum _k(Zc_k)(1)X_k,\qquad X=\sum _kc_k(1)X_k
$$
the right side is equal to
$$
\rho (X)d\pi _0(Z)\psi + \rho([Z,X])\psi.
$$
\end{proof}
\begin{remark} For $Z$ such that $\exp tZ$
represents time translation, $H$ is the energy operator,
and so is positive
in the supersymmetric model.
\end{remark}

\medskip
We shall now show that one may replace $C^\infty (\pi _0)$ by
a more or less arbitrary domain
without changing the content of the definition. This shows that
the concept of a UR of a SLG is a viable one even in
the infinite dimensional context.

\medskip
 Let us consider a system $(\pi _0, \rho, {\bb}, {\hh})$ with the following
properties.
\begin{itemize}
\item [(a)] ${\bb}$ is a dense super linear subspace of ${\hh}$ 
invariant under $\pi _0$ and ${\bb}\subset D(d\pi_0 (Z))$ for all $Z\in
[{\frak g}_1,{\frak g}_1]$;
\item [(b)] $(\rho(X))_{X\in {\frak g}_1}$ is a set of linear operators
such that:
\begin{itemize}
\item [(i)] $\rho (X)$ is symmetric for all $X\in \frak g_1$,
\item [(ii)] ${\bb}\subset D(\rho(X))$ for all $X\in \frak g_1$,
\item [(iii)] $\rho(X){\bb}_i \subset {\hh}_{i+1\,({\rm mod}\, 2)}$
for all $X\in \frak g_1$,
\item [(iv)] $\rho (aX+bY) = a \rho (X) + b \rho (Y)$ on ${\bb}$ for $X,Y\in \mathfrak{g}_1$
and $a,b$ scalars,
\item [(v)] $\pi _0(g)\rho (X)\pi _0(g)^{-1}=\rho (gX)$ on ${\bb}$
for all $g\in G_0, X\in \frak g_1$,
\item [(vi)] $\rho(X){\bb}\subset D(\rho(Y))$ for all $X,Y\in \frak g_1$,
and $-id\pi _0([X,Y])=\rho (X)\rho (Y)+\rho (Y)\rho (X)$ on ${\bb}$.
\end{itemize}
\end{itemize}
\begin{proposition}
\label{p1.3.3}
 Let $(\pi _0, \rho, {\bb}, {\hh})$ be
as above. We then have the following.
\begin{itemize}
\item [{\rm (}a{\rm )}] For any $X\in \mathfrak g_1$, $\rho (X)$
is
essentially self adjoint and $C^\infty (\pi _0)\subset D(\overline {\rho
(X)})$.

\smallskip \item [{\rm (}b{\rm )}] Let
$\overline \rho (X)=\overline {\rho (X)}\big|_{C^\infty (\pi _0)}$
for $X\in \mathfrak g_1$. Then $(\pi _0, \overline \rho, {\hh})$ is a
{\rm UR} of the {\rm SLG} $(G_0, \mathfrak g)$.
\end{itemize}
If $(\pi _0, \rho^\prime , {\hh})$ is a {\rm UR} of the {\rm SLG} $(G_0,
\mathfrak g)$,
such that
${\bb}\subset D(\overline {\rho ^\prime (X)})$ and $\overline
{\rho ^\prime (X)}$ restricts to
$\rho (X)$ on ${\bb}$ for all $X\in \mathfrak g_1$, then $\rho ^\prime =
\overline \rho $.
\end{proposition}

\begin{proof} Let $X\in \mathfrak g_1$. By assumption ${\bb}$ is invariant
under
the one parameter unitary
group generated by $H=-(1/2)id\pi _0([X,X])$ while $H=\rho (X)^2$ on ${\bb}$.
So, by Lemma \ref{l1.2.1},
$\rho (X)$ is essentially self adjoint, $\overline {\rho (X)}
= \overline {\left. \rho (X)\right|_{\bb}}$, $H=(\overline {\rho (X)})^2$, and
$D(H)\subset D(\overline {\rho (X)})$. Since $C^\infty (\pi _0)\subset
D(H)$, we
have proved (a).

\medskip
Let us now prove (b). If $a$ is scalar and $X\in \frak g$, $\overline {\rho }(a X)
= a \overline {\rho }(X)$ follows from item (iv)
and the fact that $\overline {\rho (X)}
= \overline {\left. \rho (X)\right|_{\bb}}$.  For the additivity of $\overline \rho$, let $X,Y\in
\mathfrak{g}_1$.
Then $\rho (X+Y)$ is essentially self adjoint and
its closure restricts to $\overline \rho (X+Y)$ on $C^\infty (\pi _0)$.
Then, viewing $\overline {\rho (X)}+\overline {\rho (Y)}$ as
a symmetric operator defined on the intersection of the domains
of the two operators (which includes $C^\infty (\pi _0)$),
we see that $\overline {\rho (X)}+\overline {\rho (Y)}$ is a symmetric
extension
of $\left. \rho (X) \right|_{\bb} +\left. \rho (Y) \right|_{\bb} =
\left. \rho (X+Y) \right|_{\bb}$. But as $\left. \rho (X+Y) \right|_{\bb}$
is essentially self adjoint, we have, by the remark made earlier,
$$
\overline {\rho (X)}+\overline {\rho (Y)}\prec \overline {\left. \rho (X+Y) \right|_{\bb}}
= \overline {\rho (X+Y)}.
$$
Restricting both of these operators to $C^\infty (\pi _0)$ we find
that
$\overline \rho (X+Y)=\overline \rho (X)+\overline \rho (Y)$.
From the relation $\pi _0(g)\rho (X)\pi _0(g)^{-1}=\rho (gX)$ on ${\bb}$
follows $\pi _0(g)\overline\rho (X)\pi _0(g)^{-1}=\overline\rho (gX)$.

\medskip
The key step is now to prove that for any $X\in \mathfrak g_1$, $\overline
{\rho (X)}$ maps
$C^\infty (\pi _0)$ into itself. Fix $X\in \mathfrak g_1, \psi \in C^\infty
(\pi _0)$. Now
$$
\pi _0(g)\overline {\rho (X)}\psi =\overline {\rho (gX)}\pi _0(g)\psi .
$$
So, writing $gX=\sum
_kc_k(g)X_k$ where the $c_k$ are smooth functions on $G_0$ and the $(X_k)$
a basis for
$\mathfrak g_1$, we have, remembering the linearity of
$\overline \rho $ on $C^\infty (\pi _0)$,
$$
\pi_0 (g)\overline {\rho (X)}\psi =\sum _kc_k(g)\overline{\rho (X_k)}\pi_0
(g)\psi .
$$
It is thus enough to show that
$g\longmapsto\overline {\rho (X_k)}\pi_0 (g)\psi $ is
smooth. If $H_k=-[X_k, X_k]/2$, we know from
Lemma \ref{l1.3.1}  that $\pi _0(g)\psi $ lies in $D(H_k)$
and $id\pi _0(H_k)\pi _0(g)\psi =\overline{\rho (X_k)}^2\pi _0(g)\psi $ is smooth
in $g$. Lemma \ref{l1.2.2} now shows that $\overline {\rho (X_k)}\pi _0(g)\psi $
is
smooth in $g$.

\medskip
It remains only to show that for $X, Y\in \mathfrak g_1$ we have
$$
-id\pi _0([X,Y])=\overline \rho (X)\overline \rho (Y)+
\overline \rho (Y)\overline \rho (X)
$$on $C^\infty (\pi
_0)$.
But, the right side is $\overline \rho (X+Y)^2-
\overline \rho (X)^2-\overline \rho (Y)^2$ while the left side is the
restriction of $(-i/2) d\pi _0([X+Y,X+Y])+(i/2) d\pi _0([X,X])+(i/2) d\pi _0([Y,Y])$ to
$C^\infty (\pi _0)$, and so we are done.

\medskip
We must show the uniqueness of $\overline \rho$. Let $\rho^\prime$
have the required properties also. Then $\rho^\prime (X)$
is essentially self adjoint
on $C^\infty (\pi _0)$ and ${\bb}$ is a core for its closure, by Lemma \ref{l1.2.1}. Hence
$\rho^\prime (X)=\overline \rho (X)$. The proof is complete.
\end{proof}

\medskip
We shall now prove a variant of the above result involving analytic
vectors.

\begin{proposition}
\label{p1.3.4} {\rm (}i{\rm )} If $(\pi_0,\rho,\hh)$ is a {\rm UR} of the {\rm SLG} $(G_0,\mathfrak g)$, 
then $\rho(X)$ maps
$C^\omega (\pi _0)$ into itself for all $X\in \mathfrak g_1$, so that $\pi$, as in Proposition (\ref{p1.3.2}), 
is a representation of $\mathfrak g$ in $C^\infty\left(\pi_0 \right)$.
{\rm (}ii{\rm )} Let $G_0$ be connected.
Let $\pi _0$ be an even unitary representation
of $G_0$ and ${\bb}\subset C^\omega (\pi _0)$ a dense linear super
subspace.
Let $\pi $ be a representation of $\mathfrak g$ in ${\bb}$
such that $\pi (Z)\prec d\pi _0(Z)$ for $Z\in \mathfrak g_0$ and
$\rho (X)=\zeta \pi (X)$ is symmetric for $X\in \mathfrak g_1$. Then, for each
$X\in \mathfrak g_1$, $\rho (X)$ is essentially self adjoint on ${\bb}$
and $C^\infty (\pi _0)\subset D(\overline {\rho (X)})$. If $\overline \rho
(X)$
is the restriction of $\overline {\rho (X)}$ to $C^\infty (\pi _0)$, then
$(\pi _0, \overline \rho , {\hh})$ is a {\rm UR} of the {\rm SLG} $(G_0,
\mathfrak g)$
and is the unique one in the following sense:
if $(\pi _0, \rho ^\prime , {\hh})$ is a {\rm UR} with ${\bb} \subset
D(\overline {\rho ^\prime (X)})$ and ${\rho ^\prime
(X)}\left.\right|_{\bb}= \rho(X)$  for all $X\in \mathfrak g_1$, then
$\rho ^\prime =\overline \rho$.
\end{proposition}
\begin{proof}
{\rm (}i{\rm )} This is proved as its $C^\infty$ analogue in the proof of Proposition \ref{p1.3.3},
using the analytic parts of Lemmas \ref{l1.2.2} and \ref{l1.3.1}.

\medskip
{\rm (}ii{\rm )} The proof that $\rho (X)$ for $X\in \mathfrak
g_1$ is essentially self adjoint with $D(\overline {\rho
(X)})\supset C^\infty (\pi _0)$ follows as before from (the
analytic part of) Lemma \ref{l1.2.1}. The same goes for the
linearity of $\overline \rho$.

We shall now show that for $X\in \mathfrak g_1, g\in G_0$,
\begin{equation}\label{(1)}
\pi _0(g^{-1})\overline {\rho (X)}\pi _0(g)=\overline {\rho
(g^{-1}X)}.
\end{equation}
Write
$g^{-1}X=\sum _kc_k(g)X_k$ where $(X_k)$ is a basis for $\mathfrak g_1$
and the $c_k$ are analytic functions on $G_0$.
We begin by showing that for all $\psi \in {\bb}$
\begin{equation}\label{(2)}
\overline {\rho (X)}\pi _0(g)\psi =\pi _0(g){\rho (g^{-1}X)}\psi.
\end{equation}
Now
$$
\pi _0(g) {\rho (g^{-1}X)}\psi =\sum _kc_k(g)\pi _0(g)\rho
(X_k)\psi .
$$
We argue as in Proposition \ref{p1.3.3}
to conclude, using Lemmas \ref{l1.2.2} and \ref{l1.3.1}, that the function $\overline {\rho (X)}\pi
_0(g)\psi $ is
analytic in $g$ and its derivatives can be calculated explicitly.
It is also clear that the right side is analytic in $g$ since $\rho
(X_k)\psi
\in {\bb}$ for all $k$. So, as $G_0$
is connected, it is enough to prove that the two sides in (\ref{(2)}) have all
derivatives equal at $g=1$. This comes down to showing that
for any integer $n\ge 0$ and any $Z\in \mathfrak g_0$,
\begin{equation}\label{(3)}
\rho (X)d\pi _0(Z)^n\psi =\sum _{k,r}{n\choose r}(Z^rc_k)(1)
d\pi _0(Z)^{n-r}\rho (X_k)\psi.
\end{equation}
Let $\lambda $ be the representation of $G_0$ on $\mathfrak g_1$ and write
$\lambda $ again for $d\lambda $. Then,
taking $g_t=\exp (tZ)$,
$$
\lambda (g_t^{-1})(X)=\sum _kc_k(g_t)X_k,
$$
from which we get, on differentiating $n$ times with respect to $t$ at
$t=0$,
$$
\lambda (-Z)^r(X)=\sum _k(Z^rc_k)(1)X_k.
$$
Hence the right side of (\ref{(3)}) becomes
$$
\sum _r{n\choose r}d\pi _0(Z)^{n-r}\rho (\lambda (-Z)^r(X))\psi .
$$
On the other hand, from the fact that $\pi$ is a representation of
$\mathfrak g$ in ${\bb}$ we get
$$
\rho (X)d\pi _0(Z)=d\pi _0(Z)\rho (X)+\rho (\lambda (-Z)(X))
$$
on ${\bb}$. Iterating this we get, on ${\bb}$,
$$
\rho (X)d\pi _0(Z)^n=\sum _r{n\choose r}d\pi_0(Z)^{n-r}\rho (\lambda (-Z)^r(X))
$$
which gives (\ref{(2)}). But then (\ref{(1)}) follows from (\ref{(2)}) by a simple closure
argument.

\medskip
Using (\ref{(1)}), the proof that $\overline \rho (X)$ maps $C^\infty (\pi _0)$ into itself is the same of Proposition \ref{p1.3.3}. The proof of the relation $-id\pi _0([X,Y])=\overline \rho (X)\overline \rho (Y)+
\overline \rho (Y)\overline \rho (X)$
for $X,Y \in \mathfrak g _1$ is also the same. The proof is complete.
\end{proof}

\subsection{The category of unitary representations of a super Lie group}
\label{ssec:1.4}
If $\Pi=(\pi _0, \rho, {\hh})$ and $\Pi ^\prime=(\pi ^\prime_0,
\rho ^\prime, {\hh}^\prime)$ are two UR's of a SLG $(G_0, \mathfrak g)$, a
{\em morphism} $A:\Pi \longrightarrow \Pi ^\prime $ is
an even bounded linear operator
from ${\hh}$ to ${\hh}^\prime$ such that $A$ intertwines $\pi _0, \rho $
and $\pi _0^\prime,
\rho ^\prime$; notice that as soon as $A$ intertwines $\pi _0$ and
$\pi _0^\prime$, it
maps $C^\infty (\pi _0)$ into $C^\infty (\pi _0^\prime )$, and so the
requirement that it intertwine $\rho $ and $\rho ^\prime $ makes sense. An
{\em isomorphism} is then a morphism $A$ such that
$A^{-1}$ is a bounded operator; in
this case $A$ is a linear isomorphism of $C^\infty (\pi _0)$ with
$C^\infty (\pi _0^\prime )$ intertwining $\rho $ and $\rho ^\prime$.
If $A$ is unitary we then speak of unitary equivalence of $\Pi $ and
$\Pi ^\prime $. It is easily checked that equivalence implies unitary
equivalence, just as in the classical case.
$\Pi ^\prime $ is a {\em subrepresentation} of
$\Pi$ if ${\hh}^\prime $ is a closed graded subspace of ${\hh}$ invariant under
$\pi _0$ and $\rho$, and $\pi ^\prime _0$ (resp. $\rho ^\prime $) is the restriction
of $\pi _0$
(resp. $\rho $) to ${\hh}^\prime $ (resp. $C^\infty (\pi
_0)\cap{\hh}^\prime$).
The UR $\Pi $ is said to be {\it irreducible\/} if there is no proper
nonzero
closed graded subspace ${\hh}^\prime $ that defines a subrepresentation. If
$\Pi ^\prime $ is a nonzero proper subrepresentation of $\Pi$,
and ${\hh}^{\prime \prime }$
is ${{\hh}^\prime}^\perp$,
it follows from the self adjointness of $\rho (X)$ for $X\in \mathfrak g_1$
that ${\hh}^{\prime \prime }\cap C^\infty (\pi _0)$ is invariant
under all $\rho (X) (X\in \mathfrak g_1)$;
then the restrictions of $\pi _0, \rho $ to
${\hh}^{\prime \prime }$ define a subrepresentation $\Pi ^{\prime \prime
}$ such
that $\Pi =\Pi ^\prime \oplus \Pi ^{\prime \prime }$ in an obvious manner.
\begin{lemma}
\label{l1.4.1}
$\Pi $ is irreducible if and only if ${\rm Hom }(\Pi
,\Pi )
={\complex}$.
\end{lemma}
\begin{proof}
 If $\Pi $ splits as above, then the orthogonal projection
${\hh}\longrightarrow {\hh}^\prime $ is a nonscalar element of ${\rm
Hom}(\Pi , \Pi )$.
Conversely, suppose that $\Pi $ is irreducible and $A\in {\rm Hom}(\Pi ,
\Pi )$. Then
$A^\ast \in {\rm Hom}(\Pi , \Pi )$ also and so, to prove that $A$ is a
scalar we may
suppose that $A$ is self adjoint. Let $P$ be the spectral measure of $A$.
Clearly all the $P(E)$ are even. Then $P$ commutes with $\pi _0$ and so
$P(E)$ leaves $C^\infty (\pi _0)$ invariant for all Borel sets $E$.
Moreover, by Lemma \ref{l1.2.3}, the relation $A\rho (X)=\rho (X)A$ on
$C^\infty (\pi _0)$ implies that $P(E)\leftrightarrow
\overline {\rho (X)}$ for all $E$ and $x\in \mathfrak g_1$, and
hence that $P(E)\rho (X)=\rho (X)P(E)$ on $C^\infty (\pi _0)$
for all $E, X$. The range of $P(E)$ thus defines a subrepresentation
and so $P(E)=0$ or $I$. Since this is true for all $E$, $A$
must be a scalar.
\end{proof}
\begin{lemma}
\label{l1.4.2}
 Let $(R_0, \mathfrak r)$ be a {\rm SLG} and
$(\theta , \rho^\theta)$ a {\rm UR} of it, in a Hilbert
space ${\ll}$. Let $P^X$ be the spectral
measure of $\overline {\rho^\theta (X)}, X\in \mathfrak r_1$.
Then the following
properties of a closed linear subspace ${\mm}$ of ${\ll}$ are
equivalent: {\rm (}i{\rm )} ${\mm}$ is stable under $\theta$ and
${\mm}^\infty =C^\infty (\theta )\cap {\mm}$
is stable under all $\rho^\theta (X),
(X\in \mathfrak r_1)$ {\rm (}ii{\rm )} ${\mm}$ is stable
under $\theta$
and all the spectral projections $P^X_F$ (Borel $F\subset {\real}$).
In particular, $(\theta , \rho^\theta)$ is irreducible if and
only if ${\ll}$ is irreducible under $\theta $ and all $P^X_F$.
\end{lemma}
\begin{proof} Follows from Lemma \ref{l1.2.3} applied to the orthogonal
projection $L:{\ll}\longrightarrow {\mm}$. Indeed, suppose that
${\mm}$ is a closed linear subspace of ${\ll}$ stable under
$\theta$. Then $L$ maps ${\ll}^\infty $ onto ${\mm}^\infty$.
By Lemma \ref{l1.2.3} $L$ commutes with $\rho^\theta (X)$ on
${\ll}^\infty$ if and only if $L\leftrightarrow \overline
{\rho^\theta (X)}$; this is the same as saying that
$P^X$ stabilizes ${\mm}$.
\end{proof}

\section{ Induced representations of
super Lie groups, super systems of imprimitivity, and
the super imprimitivity theorem}
\label{sec:2}

\subsection{Smooth structure of the classical induced representation
and its system of imprimitivity}
\label{ssec:2.1}
Let $G_0$ be a unimodular Lie group and $H_0$ a closed Lie subgroup.
We write $\Omega =G_0/H_0$ and assume that
$\Omega $ has a
$G_0$-invariant measure; one can easily modify the treatment below to
avoid
these assumptions. We write $x\mapsto \overline x$ for the natural map
from $G_0$ to $\Omega $ and $d\overline x$ for
a choice of the invariant measure on $\Omega $.
For any UR
$\sigma $ of $H_0$ in a Hilbert space ${\kk}$ one has the representation
$\pi $ of $G_0$ induced by $\sigma $. One may take $\pi $ as acting
in the Hilbert space ${\hh}$ of (equivalence classes) of Borel functions
$f$ from  $G_0$ to ${\kk}$ such that (i) $f(x\xi )=\sigma (\xi )^{-1}f(x)$
for all  $x\in G_0, \xi \in H_0$, and (ii) $||f||_{\hh}^2:=
\int _\Omega|f(x)|_{\kk}^2d\overline x<\infty $. Here $|f(x)|_{\kk}$ is the
norm
of $f(x)$ in $\cal K$, and the function $x\mapsto |f(x)|_{\kk}^2$ is
defined
on $\Omega $ so that it makes sense to integrate it on $\Omega $.
Let $P$ be the natural projection valued measure on ${\hh}$ defined as
follows:
for any Borel set $E\subset \Omega $ the projection $P(E)$ is the operator
$f\mapsto \chi _Ef$ where $\chi _E$ is the characteristic function of $E$.
Then
$(\pi , {\hh}, P)$ is the classical system of imprimitivity (SI)
associated
to the UR $\sigma $ of $H_0$.
In our case $G_0$
is a Lie group and it is better to work with a smooth version of
$\pi$; its structure is determined by a well known theorem
of Dixmier-Malliavin in a manner that will be explained below.

\medskip
We begin with a standard but technical lemma that says that
certain integrals containing a parameter are smooth.
\begin{lemma}
\label{l2.1.1}
 Let $M, N$ be smooth manifolds, $dn$
a smooth measure on $N$, and $B$ a separable Banach space with
norm $|{\cdot}|$. Let
$F:M\times N\longrightarrow B$ be a map with the following properties:
{\rm (}i{\rm )} For each $n\in N$, $m\mapsto F(m,n)$ is smooth
{\rm (}ii{\rm )} If $A\subset M$ is an open set with compact closure,
and $G$ is any derivative
of $F$ with respect to $m$, there is a
$g_A\in L^1(N, dn)$ such that $|G(m,n)|\le g_A(n)$ for all
$m\in A, n\in N$. Then
$$
f(m)=\int _NF(m,n)dn
$$
exists for all $m$ and $f$ is a smooth map of $M$ into $B$.
\end{lemma}
\begin{proof}
 It is a question of proving that the integrals
$$
\int _N|G(m,n)|dn
$$
converge uniformly when $m$ varies in an open subset $A$ of $M$
with compact closure. But the integrand is majorized by
$g_A$ which is integrable on $N$
and so the uniform convergence is clear.
\end{proof}

\medskip
We also observe that any $f\in {\hh}$ lies in
$L^{p, {\rm loc}}(G_0)$ for $p=1,2$, i.e., $\theta (x)|f(x)|_{\kk}^2$
is integrable on $G_0$ for any continuous compactly supported
scalar function $\theta \ge 0$. In fact
$$
\int _{G_0}\theta (x)|f(x)|^2_{\kk} dx=\int _\Omega\left (\int _{H_0}\theta
(x\xi )|f(x\xi )|_{\kk}^2 d\xi\right )d\overline x=
\int _\Omega\overline \theta (x)|f(x)|_{\kk}^2d\overline x<\infty
$$
where $\overline \theta (x)=\int _{G_0}\theta (x\xi )d\xi$.

\medskip
In ${\hh}$ we have the space $C^\infty (\pi )$ of smooth vectors
for $\pi$. We also have its Garding  subspace, the subspace spanned
by all vectors $\pi (\alpha )h$ where $\alpha \in C_c^\infty (G_0)$
and $h\in {\hh}$. We have
$$
(\pi (\alpha )h)(z)=\int _{G_0}\alpha (x)h(x^{-1}z)dx
=\int _{G_0}\alpha (zt^{-1})h(t)dt\qquad (z\in G_0).
$$
The integrals exist because $h$ is locally $L^2$ on $G_0$ as
mentioned above.
Since $h\in L^{1, {\rm loc}}(G_0), \alpha \in C_c^\infty (G_0)$,
the conditions of Lemma \ref{l2.1.1} are met and so $\pi (\alpha )h$
is smooth. Thus all elements of the Garding space are smooth
functions. But
the Dixmier-Malliavin theorem asserts that the Garding space is exactly
the same as $C^\infty (\pi )$ \cite{DIM}.
Thus all elements of $C^\infty (\pi )$
are smooth functions from $G_0$ to ${\kk}$. This is the key point
that leads to the smooth versions of the induced representation
and the SI at the classical level.

\medskip
Let
us define ${\bb}$ as the space of all functions $f$ from $G_0$ to ${\kk}$
such that (i) $f$ is smooth and $f(x\xi )=\sigma (\xi )^{-1}f(x)$ for
all $x\in G_0, \xi \in H_0$ (ii) $f$ has compact support mod $H_0$.
Let $C_c^\infty (\pi )$ be the subspace of all elements of $C^\infty (\pi
)$
with compact support mod $H_0$.
\begin{proposition}
\label{p2.1.2}
${\bb}$ has the following properties:
{\rm (}i{\rm )}
${\bb}=C_c^\infty (\pi )$ {\rm (}ii{\rm )} $\bb$ is dense in $\hh$
{\rm (}iii{\rm )} $f(x)\in C^\infty
(\sigma )$ for all $x\in G_0$
{\rm (}iv{\rm )} $\bb$ is stable under $d\pi$.
\end{proposition}
\begin{proof} (i) Let $f\in {\bb}$. To show that $f\in C^\infty (\pi )$
it is enough to show that for any $u\in {\hh}$
the map $x\mapsto (\pi (x^{-1})f,u)_{\hh}$ is smooth in $x$. Now
$$
(\pi (x^{-1})f,u)_{\hh}=\int _\Omega (f(xy),u(y))_{\kk}d\overline y.
$$
Since $|u|_{\kk}$ is locally $L^1$ on $X$ and $f$ is smooth,
the conditions of Lemma \ref{l2.1.1} are met. We
have ${\bb}\subset C_c^\infty (\pi )$. The reverse
inclusion is immediate from
the Dixmier-Malliavin theorem, as remarked above.
\medskip\\
(ii) It is enough to prove that any $h\in {\hh}$ with compact support mod
$H_0$
is in the
closure of $\bb$. We know that $\pi (\alpha )h\to h$ as $\alpha
\in C_c^\infty (G_0)$ goes
suitably to the delta function at the identity of $G_0$. But $\pi (\alpha
)h$
is smooth and has compact support mod $H_0$ because $h$ has the same
property,
so that $\pi (\alpha )h\in {\bb}$.
\medskip\\
(iii) Fix $x\in G_0$. Since $\sigma (\xi )f(x)=f(x\xi ^{-1})$ for $\xi \in
H_0$
it is clear that $f(x)\in C^\infty (\sigma)$.
\medskip\\
(iv) Let $f\in {\bb}, Z\in \mathfrak g_0$. Then
$$
(d\pi (Z)f)(x)=(d/dt)_{t=0}f(\exp (-tZ)x)
$$
is smooth and we are done.
\end{proof}

\medskip
We refer to $(\pi ,{\bb})$ as the {\it smooth representation
induced by\/} $\sigma $. We shall also define the smooth version
of the SI. For any $u\in C_c^\infty (\Omega )$ let $M(u)$ be the
bounded operator on ${\hh}$ which is multiplication by $u$. Then
$M(u)$ leaves ${\bb}$ invariant and $M:u\mapsto M(u)$ is a
$\ast$-representation of the $\ast$-algebra $C_c^\infty (\Omega )$
in ${\hh}$. It is natural to refer to $(\pi , {\bb}, M)$ as the
{\em smooth system of imprimitivity} associated to $\sigma$. Observe that $f\in C^\infty (\pi
)$ has compact support mod $H_0$ if and only if there is some
$u\in C_c^\infty (\Omega )$ such that $f=M(u)f$. Proposition
\ref{p2.1.2} shows that ${\bb}$ is thus determined intrinsically
by the SI associated to $\sigma $. The passage from $(\pi , {\hh},
P)$ to $(\pi , {\bb}, M)$ is thus functorial and is a categorical
equivalence. Thus we are justified in working just with smooth
SI's.

\medskip
It is easy to see that the assignment that takes $\sigma $
to the associated
smooth SI is functorial. Indeed, let $R$ be a morphism from
$\sigma $ to $\sigma '$, i.e., $R$ is a bounded operator from
${\kk}$ to ${\kk}'$ intertwining $\sigma $ and $\sigma '$.
We then define $T_R=T({\hh}\longrightarrow
{\hh}')$
by $(T_Rf)(x)=Rf(x) (x\in G_0)$. It is then immediate that
$T_R$ is a morphism
from the (smooth) SI associated to $\sigma $ to the
(smooth) SI associated to
$\sigma '$. This functor is an equivalence of categories.
To
verify this one must show that every morphism between the two
SI's is of this form.
This is of course classical but we sketch the argument
depending on the following lemma which
will be essentially used in the super context also.

\begin{lemma}
\label{l2.1.3}
Suppose $f\in {\bb}$ and $f(1)=0$. Then
we can find $u_i\in C_c^\infty (\Omega ), g_i\in {\bb}$ such that
{\rm (}i{\rm )}
$u_i(1)=0$ for all $i$ {\rm (}ii{\rm )}
we have $f=\sum _iu_ig_i$.
\end{lemma}
\begin{proof}
 If $f$ vanishes in a neighborhood of $1$, we can choose
$u\in C_c^\infty (\Omega)$ such that $u=0$ in a neighborhood of $1$
and $f=uf$. The result is thus true for $f$. Let
$f\in {\bb}$ be arbitrary but vanishing at $1$. Let $\mathfrak z$ be a linear
subspace of $\mathfrak g_0=$ Lie $(G_0)$ complementary to $\mathfrak h_0=$
Lie $(H_0)$. Then there is a sufficiently small $r>0$ such that if
$\mathfrak z_r=\{Z\in \mathfrak z\ |\ |Z|<r\}$, $|{\cdot}|$ being a norm
on $\mathfrak z$, the map
$$
\mathfrak z_r\times H_0\longrightarrow G_0,\qquad (Z, \xi )
\longmapsto \exp Z{\cdot}\xi
$$
is a diffeomorphism onto an open set $G_1=G_1H_0$ of $G_0$. We
transfer $f$
from $G_1$ to
a function, denoted by $\varphi$ on $\mathfrak z_r\times H_0$. We have
$\varphi(0,\xi )=0$,
and $\varphi (Z, \xi \xi ')=\sigma (\xi ')^{-1}\varphi (Z, \xi )$
for $\xi '\in H_0$. If $t_i (1\le i\le k)$ are the linear
coordinates on $\mathfrak z$,
$$
\varphi (Z, \xi )=\sum _i t_i(Z)\int _0^1(\partial
\varphi/\partial t_i)(sZ, \xi )ds.
$$
The functions $\psi _i(Z, \xi )=\int _0^1(\partial \varphi/\partial
t_i)(sZ, \xi )ds$
are smooth by Lemma \ref{l2.1.1} while $\psi _i (Z, \xi \xi ')=
\sigma (\xi ')^{-1}\psi _i (Z, \xi )$
for $\xi '\in H_0$.
So, going back to $G_1$ we can write
$f=\sum _it_ih_i$
where $t_i$ are now in $C^\infty (G_1)$, right invariant
under $H_0$ and
vanishing at $1$, while the $h_i$ are smooth and
satisfy $h_i(x\xi )=
\sigma (\xi )^{-1}h_i(x)$ for $x\in G_1, \xi \in H_0$.
If $u\in C_c^\infty (\Omega )$
is such that $u$ is $1$ in a neighborhood of $1$ and
supp $(u)\subset G_1$, then
$u^2f=\sum _iu_ig_i$ where $u_i=ut_i\in C_c^\infty (\Omega ),
u_i(1)=0$, and $g_i=uh_i\in {\bb}$.
Since $f=u^2f+(1-u^2)f$ and $(1-u^2)f=0$ in a neighborhood
of $1$, we are done.
\end{proof}

\medskip
We can now determine all the morphisms from ${\hh}$ to ${\hh}'$.
Let $T$
be a morphism ${\hh}\longrightarrow {\hh}'$. Then,  as $T$ commutes with
multiplications by elements
of $C_c^\infty (\Omega )$, it maps
${\bb}$ to ${\bb}'$.
Moreover, for the same reason,
the above lemma shows that if $f\in {\bb}$ and $f(1)=0$,
then $(Tf)(1)=0$. So the map
$$
R : f(1)\longmapsto (Tf)(1)\qquad (f\in {\bb})
$$
is well defined. From the fact that $T$ intertwines $\pi $ and $\pi '$
we obtain that $(Tf)(x)=Rf(x)$ for all $x\in G_0$. To complete
the proof we must show two things: (1) $R$ is defined on all of
$C^\infty (\sigma )$ and (2) $R$ is bounded. For (1), let
$v\in C^\infty (\sigma )$. In the earlier notation, if
$u\in C_c^\infty (G_0)$ is $1$ in 1 and has support contained in $G_1$, then $h:(\exp Z, \xi )\mapsto u(\exp Z)
\sigma (\xi )^{-1}v$ is in ${\bb}$ and $h(1)=v$. For proving (2),
let the constant $C>0$ be such that
$$
||Tg||_{{\hh}'}\le C||g||_{\hh}\qquad (g\in {\hh}').
$$
Then, taking $g=u^{1/2}f$ for $f\in {\bb}$ and $u\ge 0$
in $C_c^\infty (\Omega )$, we get
$$
\int _\Omega u(x)|Rf(x)|_{\kk}^2 d\overline x\le C
\int _\Omega u(x)|f(x)|_{\kk'}^2 d\overline x
$$
for all $f\in {\bb}$ and $u\ge 0$ in $C_c^\infty (\Omega )$. So
$|Rf(x)|_{\kk}\le C |f(x)|_{\kk}$ for almost all $x$. As
$f$ and $Rf=Tf$ are continuous this inequality is valid for all $x$,
in particular for $x=1$, proving that $R$ is bounded.

\bigskip\noindent
\subsection{Representations induced from a special sub super Lie group}
\label{ssec:2.2}
It is now our purpose to extend this smooth classical
theory to the super context.
A SLG $(H_0, \mathfrak h)$ is a {\em sub super Lie group}  of the SLG $(G_0, \mathfrak g)$ if $H_0
\subset G_0, \mathfrak h\subset \mathfrak g$, and the action of $H_0$ on $\mathfrak h$
is the restriction of the action of $H_0$ (as a subgroup of $G_0$) on
$\mathfrak g$. We shall always suppose that $H_0$ is
closed in $G_0$. The sub SLG $(H_0, \mathfrak h)$ is called
{\it special\/} if $\mathfrak h$
has the same odd
part as $\mathfrak g$, i.e., $\mathfrak h_1=\mathfrak g_1$. In this case the super
homogeneous space associated is purely even and coincides with
$\Omega =G_0/H_0$. As in \S\ref{ssec:2.1} we shall assume that $\Omega $ admits an
invariant measure although it is not difficult to modify
the treatment to avoid this assumption.
Both conditions are satisfied in the case of the super Poincar\'e groups
and their variants.

\medskip
We start with a UR
$(\sigma ,\rho ^\sigma , {\kk})$ of $(H_0, \mathfrak h)$
and associate to it the smooth
induced representation $(\pi , {\bb})$ of the classical
group $G_0$. In our case $\cal K$ is a SHS
and so $\cal H$ becomes a SHS
in a natural manner,
the parity subspaces being the subspaces where $f$ takes its values in
the corresponding parity subspace of $\kk$. $\pi $ is an even UR.

\medskip
We shall now define the operators $\rho ^\pi (X)$ for $X\in \mathfrak g_1$
as follows:
$$
(\rho ^\pi(X)f)(x)=\rho ^\sigma (x^{-1}X)f(x)\qquad (f\in {\bb}).
$$
Since the values of $f$ are in $C^\infty (\sigma)$ the right side is well
defined. In order to prove that the definition gives us an odd operator
on ${\bb}$ we need a lemma.
\begin{lemma}
\label{l2.2.1}
 $[\mathfrak g_1,\mathfrak g_1]\subset \mathfrak h_0$ and
is stable under $G_0$. In particular it is an ideal in
$\mathfrak g_0$.
\end{lemma}
\begin{proof} For $g\in G_0, Y,Y'\in \mathfrak g_1$, we have
$g[Y,Y']=[gY,gY']
\in [\mathfrak g_1,\mathfrak g_1]$. Since $\mathfrak h_0\oplus \mathfrak g_1$ is a super
Lie
algebra, $[\mathfrak g_1,\mathfrak g_1]\subset \mathfrak h_0$.
\end{proof}
\begin{proposition}
\label{p2.2.2}
  $\rho ^\pi (X)$ is an odd linear map
${\bb}
\longrightarrow {\bb}$ for all $X\in \mathfrak g_1$. Moreover $\rho ^\pi (X)$
is local, i.e.,
${\rm supp} (\rho ^\pi (X)f)\subset {\rm supp} (f)$ for $f\in {\bb}$.
Finally, if $Z\in \mathfrak h_0$, we have $-d\sigma (Z)f(g)=(Zf)(g)$.
\end{proposition}
\begin{proof}
 The support relation is trivial.
Further, for $x\in G_0, \xi \in H_0$,

\begin{eqnarray*}
(\rho ^\pi (X)f)(x\xi )&= &\rho ^\sigma (\xi ^{-1}x^{-1}X)
f(x\xi )\\
&=&
\sigma (\xi )^{-1}\rho ^\sigma (x^{-1}X)\sigma (\xi )f(x\xi )\\
&=&\sigma (\xi )^{-1} (\rho ^\pi (X)f)(x).\\
\end{eqnarray*}
It is thus a question
of proving that $g\mapsto \rho ^\sigma (g^{-1}X)f(g)$ is smooth.
If $(X_k)$ is a basis
for $\mathfrak g_1$, $g^{-1}X=\sum _kc_k(g)X_k$ where the $c_k$ are smooth
functions
and so it is enough to prove that $g\mapsto \rho ^\sigma (Y)f(g)$ is
smooth
for any $Y\in \mathfrak g, f\in {\bb}$. We use Lemma \ref{l1.2.2}. If $Z=(1/2)[Y,Y]$,
we have $\rho ^\sigma (Y)^2f(g)=-id\sigma (Z)f(g)$, and we need only show
that
$-d\sigma (Z)f(g)$ is smooth in $g$. But $Z\in \mathfrak h_0$ and $f(g\exp
tZ)=
\sigma (\exp (-tZ))f(g)$ so that $-d\sigma (Z)f(g)=(Zf)(g)$ is clearly
smooth in $g$.
Note that this argument applies to any $Z\in \mathfrak h_0$, giving the
last assertion.
\end{proof}
\begin{proposition}
\label{p2.2.3}
 $(\pi, \rho ^\sigma ,{\bb})$ is a
{\rm UR} of the {\rm SLG} $(G_0, \mathfrak g)$.
\end{proposition}
\begin{proof}
 The symmetry of $\rho ^\pi (X)$ and the relations
$\rho ^\pi (yX)=\pi (y)\rho ^\pi (X) \pi (y)^{-1}$ follow immediately
from the corresponding relations for $\rho ^\sigma $.
Suppose now that $X, Y\in \mathfrak g_1$. Then
$(\rho ^\pi (X)\rho ^\pi (Y)f)(x)=\rho ^\sigma
(x^{-1}X)\rho ^\sigma (x^{-1}Y)f(x)$. Hence

\begin{eqnarray*}
((\rho ^\pi (X)\rho ^\pi (Y)+\rho ^\pi (Y)\rho ^\pi (X))f)(x)&=&
-id\sigma (x^{-1}[X,Y])f(x)\\
 &=&i(x^{-1}[X,Y]f)(x)
(\hbox { Proposition \ref{p2.2.2} })\\
&=&i(d/dt)_{t=0}f(x(x^{-1}\exp t[X,Y]x))\\
&=&i(d/dt)_{t=0}f(\exp t[X,Y]x)\\
&=&i(d/dt)_{t=0}(\pi (\exp (-t[X,Y])f)(x)\\
&=&-i(d\pi ([X,Y])f)(x).
\end{eqnarray*}
This proves the proposition.
\end{proof}
We refer to $(\pi, \rho ^\pi , {\hh})$ as the UR of the SLG
$(G_0, \mathfrak g)$ {\it induced\/}
by the UR $(\sigma ,\rho ^\sigma , {\kk})$ of $(H_0, \mathfrak h)$,
and to $(\pi ,\rho ^\pi ,{\bb})$ as the corresponding smooth induced UR.

\medskip
Write $P$ for the natural projection valued measure in ${\hh}$
based on $\Omega$: for any Borel $E\subset \Omega$, $P(E)$ is
the operator in ${\hh}$ of multiplication by $\chi _E$, the
characteristic function of $E$.

\medskip
Recall the definition of $\leftrightarrow$ before
Lemma \ref{l1.2.3}.
\begin{proposition}
\label{p2.2.4}
 For $X\in \mathfrak g_1$, and
$u\in C_c^\infty (\Omega )$, we have
$M(u)\leftrightarrow \overline{\rho ^\pi (X)}$. Furthermore
$P(E)\leftrightarrow \overline{\rho ^\pi (X)}$ for Borel
$E\subset \Omega $.
\end{proposition}
\begin{proof}
 It is standard that a bounded operator commutes
with all $P(E)$ if and only if it commutes with all $M(u)$
for $u\in C_c^\infty (\Omega )$. It is thus enough to prove that
$M(u)\leftrightarrow \overline{\rho^\pi (X)}$ for
all $u, X$. On ${\bb}$ we have $M(u)\rho ^\pi (X)=\rho ^\pi (X)M(u)$
trivially from the definitions,
and so we are done in view of Lemma \ref{l1.2.3}.
\end{proof}
\begin{theorem}
\label{t2.2.5}
 The assignment that takes
$(\sigma ,\rho ^\sigma)$ to
$(\pi ,\rho ^\pi ,{\bb}, M)$
is a fully faithful functor.
\end{theorem}
\begin{proof}
 Let $R$ be a morphism intertwining
$(\sigma ,\rho ^\sigma )$ and $(\sigma ', \rho ^{\sigma '})$, and let
$T:{\bb}\longrightarrow {\bb}'$ be associated to $R$ such that
$(Tf)(x)=Rf(x)$.
It is then immediate that $T$ intertwines
$\rho ^\pi $ and $\rho ^{\pi '}$. Conversely, if $T$ is a
morphism between the induced systems, from the classical
discussion following Lemma \ref{l2.1.3} we know that
$(Tf)(x)=Rf(x)$ for a bounded even operator $R$
intertwining $\sigma$ and $\sigma ^\prime$. Since
$T$ intertwines $\rho^\pi$ and $\rho^{\pi^\prime}$
we conclude that $R$ must intertwine $\rho^\sigma$
and $\rho^{\sigma^\prime}$.
\end{proof}
\subsection{Super systems of imprimitivity and the super
imprimitivity theorem}
\label{ssec:2.3}
A {\it super system of imprimitivity} (SSI)
based on $\Omega $ is a collection $(\pi ,
\rho ^\pi ,{\hh}, P)$ where $(\pi , \rho ^\pi , {\hh})$ is a UR of
the SLG $(G_0, \mathfrak g)$, $(\pi ,{\hh}, P)$ is a classical system
of imprimitivity, $\pi , P$ are both even, and $\overline{\rho ^\pi
(X)}\leftrightarrow P(E)$
for all $X\in \mathfrak g_1$ and Borel $E\subset \Omega $.

\medskip
Let $(\pi, \rho^\pi , {\hh})$ be the induced representation
defined in \S\ref{ssec:2.2}  and let $P$ be the projection valued measure
introduced above. Proposition \ref{p2.2.4} shows that
$(\pi, \rho^\pi , {\hh}, P)$ is a SSI based on $\Omega$. We call this
the SSI {\it induced by $(\sigma, \rho^\sigma )$.\/}
\begin{theorem}[super imprimitivity theorem]
\label{t2.3.1}
 The assignment that takes
$(\sigma , \rho ^\sigma )$
to $(\pi , \rho ^\pi , {\hh}, P)$ is an equivalence of categories
from the category of {\rm UR }'s of the special sub
{\rm SLG} $(H_0, \mathfrak h)$
to the category of  {\rm SSI }'s based on $\Omega $.
\end{theorem}
\begin{proof}
 Let us first prove that any SSI of
the SLG $(G_0, \mathfrak g)$ is induced from a UR of the SLG $(H_0, \mathfrak h)$.
We may assume, in view of the classical imprimitivity
theorem  that $\pi $ is the representation induced by a UR $\sigma $ of
$H_0$ in ${\kk}$
and that $\pi $ acts by left translations on ${\hh}$.  By assumption
$\rho ^\pi (X)$ leaves $C^\infty
(\pi)$ invariant.  We claim that it leaves $C_c^\infty (\pi)$ also
invariant.
Indeed, let
$f\in C_c^\infty (\pi)$; then there is $u\in C_c^\infty (\Omega )$ such
that
$f=uf$. On the other hand, by Lemma \ref{l1.2.3}, $\overline {\rho^\pi (X)}M(u)=
M(u)\overline {\rho^\pi (X)}$ so that $uf\in D(\overline {\rho^\pi (X)})$
and
$\overline {\rho^\pi (X)}(uf)=u\overline {\rho^\pi (X)}f$. Since $uf=f$
this comes to $\rho^\pi (X)f=u\rho^\pi (X)f$, showing that
$\rho^\pi (X)f\in C_c^\infty (\pi)$. Thus the $\rho^\pi (X)$ leave ${\bb}$
invariant and commute with all $M(u)$ there. In other words
we may work with the smooth SSI.

\medskip
By Lemma \ref{l2.1.3} the map $f(1)\longmapsto (\rho ^\pi (X)f)(1)$ is well
defined
and so, as in \S \ref{ssec:2.1} we can define a map
$$
\rho ^\sigma (X):C^\infty (\sigma )\longrightarrow C^\infty (\sigma )
$$
by
$$
\rho ^\sigma (X)v=(\rho ^\pi (X)f)(1), \qquad f(1)=v, \quad f\in {\bb}.
$$
Then, for $f\in {\bb}, x\in G_0$,

\begin{eqnarray*}
(\rho^\pi (X)f)(x)&=&(\pi (x^{-1})\rho^\pi (X)f)(1)\\
&=&(\rho^\pi (x^{-1}X)\pi (x^{-1})f)(1)\\
&=&\rho ^\sigma (x^{-1}X)(\pi (x^{-1})f)(1)\\
&=&\rho ^\sigma (x^{-1}X)f(x).
\end{eqnarray*}
If we now prove that $(\sigma, \rho^\sigma ,{\kk})$ is a
UR of the SLG $(H_0, \mathfrak h)$, we are done. This is completely formal.

{\it Covariance with respect to $H_0$\/}: For $f\in {\bb}, \xi \in H_0$,
\begin{eqnarray*}
\rho ^\sigma (\xi X)f(1)&=&(\rho^\pi (\xi X)f)(1)\\
&=&(\pi (\xi )\rho^\pi (X)\pi (\xi ^{-1})f)(1)\\
&=&(\rho^\pi(X)\pi (\xi ^{-1})f)(\xi ^{-1})\\
&=&\sigma (\xi )(\rho^\pi (X)\pi (\xi ^{-1})f)(1)\\
&=&\sigma (\xi )\rho^\sigma (X)(\pi (\xi ^{-1})f)(1)\\
&=&\sigma (\xi )\rho^\sigma (X)\sigma (\xi )^{-1}f(1).
\end{eqnarray*}

{\it Odd commutators\/}: Let $X, Y\in \mathfrak g_1=\mathfrak h_1$ so that
$Z=[X,Y]\in \mathfrak h_0$. We have
$$
[\rho^\pi (X),\rho^\pi (Y)]f=-id\pi ([X.Y])f\eqno (\ast)
$$
for all $f\in {\bb}$. Now,
\begin{eqnarray*}
i(-d\pi (Z)f)(1)&=&i(d/dt)_{t=0}f(\exp tZ)\\
&=&i(d/dt)_{t=0}\sigma (\exp (-tZ))f(1)\\
&=&-id\sigma (Z)f(1).
\end{eqnarray*}
On the other hand,
$$
(\rho^\pi (X)\rho^\pi (Y)f)(1)=\rho^\sigma (X)\rho^\sigma (Y)f(1)
$$
so that the left side of $(\ast)$, evaluated at $1$, becomes
$$
[\rho^\sigma (X),\rho^\sigma (Y)]f(1).
$$
Thus
$$
[\rho^\sigma (X),\rho^\sigma (Y)]f(1)=-id\sigma (Z)f(1).
$$

{\it Symmetry\/}: From the symmetry of the $\rho^\pi (X)$ we have,
for all $f,g\in {\bb}, a,b\in C_c^\infty (\Omega )$,
$$
(\rho ^\pi (X)(af),b g)_{\hh}=(a f,\rho ^\pi (X)(bg))_{\hh}.
$$
This means that
$$
\int (\rho^\sigma (x^{-1}X)f(x),g(x))_{\kk}a(x)\overline {b(x)}d\overline
x=
\int (f(x),\rho^\sigma (x^{-1}X)g(x))_{\kk}a(x)\overline {b(x)}d\overline
x.
$$
Since $a$ and $b$ are arbitrary we conclude that
$$
(\rho^\sigma (x^{-1}X)f(x),g(x))_{\kk}=(f(x),\rho^\sigma
(x^{-1}X)g(x))_{\kk}
$$
for almost all $x$. All functions in sight are continuous and so
this relation is true for {\it all\/} $x$. The evaluation at $1$ gives
the symmetry of $\rho^\sigma (X)$ on $C^\infty (\sigma )$.

\medskip
This proves that $(\sigma ,\rho ^\sigma ,{\kk})$
is a UR of the SLG $(H_0, \mathfrak h)$ and that the corresponding induced
SSI is the one we started with.

\medskip
To complete the proof we must show that the set of
morphisms of the induced SSI's is in canonical bijection
with the set of morphisms of the inducing UR's of the
sub SLG in question. Let $(\pi, \rho^\pi , {\hh}, P)$ and
$(\pi', \rho^{\pi '}, {\hh}', P')$ be the SSI's induced
by $(\sigma ,\rho^\sigma )$ and $(\sigma ',\rho^{\sigma '})$
respectively. For any morphism $R$ from $(\sigma ,\rho^\sigma )$
to $(\sigma ',\rho^{\sigma '})$ let $T$ be as in Theorem \ref{t2.2.5}.
Then $T$ extends uniquely to a bounded even operator
from ${\hh}$ to ${\hh}'$, and the relations $TM(u)=M'(u)T$
for all $u\in C_c^\infty (\Omega)$ imply that $TP(E)=P'(E)T$
for all Borel $E\subset \Omega$. Hence $T$ is a morphism from
$(\pi, \rho^\pi , {\hh}, P)$ to $(\pi', \rho^{\pi '}, {\hh}', P')$.
It is clear that the assignment $R\longmapsto T$ is
functorial. To complete the proof we must show that any
morphism $T$ from $(\pi, \rho^\pi , {\hh}, P)$ to
$(\pi', \rho^{\pi '}, {\hh}', P')$ is of this form
for a unique $R$. But $T$ must take ${\bb}=C_c^\infty (\pi _0)$
to ${\bb}'=C_c^\infty (\pi '_0)$ and commute with the actions of
$C_c^\infty (\Omega )$. Hence $T$ is a morphism from
$(\pi, \rho^\pi , {\bb}, M)$ to $(\pi', \rho^{\pi '}, {\bb}', M')$.
Theorem \ref{t2.2.5} now implies that $T$ arises from a unique
morphism of $(\sigma ,\rho^\sigma )$ to $(\sigma ',\rho^{\sigma '})$.

\medskip
This finishes the proof of Theorem \ref{t2.3.1}.
\end{proof}

\section{ Representations of super semidirect products and super Poincar\'e groups}
\label{sec:3}

\subsection{Super semidirect products and their irreducible
unitary representations}
\label{ssec:3.1}
We start with a classical semidirect
product $G_0=T_0\times 'L_0$ where $T_0$ is a vector space
of finite dimension over ${\real}$, the translation group, and
$L_0$ is a closed unimodular
subgroup of ${\rm GL}(T_0)$ acting on $T_0$
naturally.
For any Lie group the corresponding gothic letter denotes
its Lie algebra.
In applications $L_0$ is usually an orthogonal group of
Minkowskian signature, or its $2$-fold cover, the
corresponding
spin group. By a {\em super  semidirect product} (SSDP) we mean
a SLG $(G_0, \mathfrak g)$ where $T_0$ acts trivially on
$\mathfrak g_1$ and $[\mathfrak g_1,\mathfrak g_1]\subset \mathfrak t_0$.
Clearly
$\mathfrak t:=\mathfrak t_0\oplus \mathfrak g_1$ is also a
super Lie algebra, and $(T_0, \mathfrak t)$ is a SLG called the
{\it super translation group.\/} For any closed subgroup
$S_0\subset L_0$, $H_0=T_0S_0$ is a closed subgroup
of $G_0$,
$\mathfrak h=\mathfrak h_0\oplus \mathfrak g_1$ is a super Lie algebra
where $\mathfrak h_0=\mathfrak t_0\oplus \mathfrak s_0$ is the Lie algebra
of $H_0$. Notice that $(H_0, \mathfrak h)$ is a {\it special\/}
sub SLG of $(G_0, \mathfrak g)$. We begin by showing that the
irreducible UR's of $(G_0, \mathfrak g)$ are in
natural bijection with the irreducible UR's of suitable special
sub SLG's of the form $(H_0, \mathfrak h)$ with the property
that the translations act as scalars. For brevity we shall write
$S=(G_0, \mathfrak g), T=(T_0, \mathfrak t)$.

\medskip
The action of $L_0$ on $T_0$ induces an action on the dual
$T_0^\ast$ of $T_0$. We assume that this action is regular, i.e.,
the orbits are all locally closed. By the well known theorem
of Effros this implies that if $Q$ is any projection valued
measure on $T_0^\ast$ such that $Q_E=0$ or the identity operator  $I$ for any invariant
Borel subset $E$ of $T_0^\ast$, then $Q$ is necessarily concentrated on a
single orbit. This is precisely the condition under which the
classical method of little groups of Frobenius-Mackey-Wigner
works. For any $\lambda \in T_0^\ast$ let $L^\lambda _0$ be the
stabilizer
of $\lambda $ in $L_0$ and let $\mathfrak g^\lambda =
\mathfrak t_0\oplus \mathfrak l^\lambda _0\oplus
\mathfrak g_1$. The SLG $(T_0L^\lambda _0, \mathfrak g^\lambda )$
will be denoted by $S^\lambda $. We shall call it the
{\it little super group\/}
at $\lambda $. It is a special sub SLG of $(G_0, \mathfrak g)$.
Two $\lambda $'s are called equivalent
if they are in the same $L_0$-orbit. If $\theta$ is a UR of the
classical group $T_0L_0$ and $O$ is an orbit in $T_0^\ast$,
its spectrum is said to be {\it in} $O$ if the spectral measure
(via the SNAG theorem) of the restriction of $\theta$ to $T_0$
is supported by $O$.

\medskip

Given $\lambda \in T_0^\ast$, a UR $(\sigma , \rho^\sigma )$ of
$S^\lambda$ is {\em $\lambda$-admissible} if $\sigma (t)=e^{i\lambda (t)}I$
for $t\in T_0$. $\lambda$ itself is called {\em admissible} if there is an 
irreducible UR which is $\lambda$-admissible.
It is
obvious that the property of being admissible is
preserved under the action of $L_0$. Let
$$
T_0^+=\biggl\{ \lambda \in T_0^\ast \ \bigg |\
\lambda \ {\rm admissible}\ \biggr \}.
$$
Then $T_0^+$ is an invariant subset of
$T_0^\ast$.
\begin{theorem}
\label{t3.1.1}
 The spectrum of every irreducible
{\rm UR} of the {\rm SLG} $(G_0, \mathfrak g)$ is in some orbit in $T_0^+$.
For each orbit
in $T_0^+$ and choice of
$\lambda $ in that orbit, the assignment that takes a
$\lambda$-admissible {\rm UR} $\gamma :=(\sigma ,
\rho^\sigma)$ of $S^\lambda $
into the {\rm UR}
$U^\gamma $ of $(G_0, \mathfrak g)$ induced by it, is a functor
which is an
equivalence of categories between the category of the
$\lambda$-admissible {\rm UR}'s of
$S^\lambda $ and the category of {\rm UR}'s of
$(G_0, \mathfrak g)$ with
their spectra in that orbit. Varying $\lambda $
in that orbit changes the functor into an
equivalent one. In particular this
functor gives a bijection between the
respective sets of equivalence classes
of irreducible {\rm UR}'s.
\end{theorem}
\begin{proof}
Notice first of all that since $T_0$ acts trivially on $\mathfrak{g}_1$, 
$\pi_0(t)$ commutes with $\rho^\pi \left( X \right)$ on $C^\infty\left(\pi_0\right)$
for all $t\in T_0,\, X\in \mathfrak{g}_1$. Hence $P_E\leftrightarrow \rho^\pi\left(X\right)$ 
for all Borel $E\subset \Omega,\,X\in \mathfrak{g}_1 $ .
 For the first statement,
let $E$ be an {\it invariant\/}
Borel subset of $T_0^\ast$. Let $P$
be the spectral measure of the restriction
of $\pi$ to $T_0$. Then $P_E$
commutes with $\pi, \rho^\pi (X)$.
So, if $(\pi , \rho^\pi)$ is irreducible, $P_E=0$ or $I$.
Hence
$P$ is concentrated in some orbit $O$, i.e., $P_O=I$.
The system
$(\pi ,\rho^\pi )$ is clearly equivalent to
$(\pi , \rho^\pi ,P)$ since
$P$ and the restriction of $\pi$ to $T_0$ generate
the same algebra. If
$\lambda \in O$ and $L^\lambda _0$ is the
stabilizer of $\lambda $ in $L_0$,
we can transfer $P$ from $O$ to a projection valued
measure $P^\ast$ on
$L_0/L^\lambda _0=T_0L_0/T_0L^\lambda _0$. So
$(\pi , \rho^\pi )$ is equivalent to the SSI
$(\pi , \rho^\pi , P^\ast)$.
The rest of the theorem is an immediate consequence
of Theorem \ref{t2.3.1}.
The fact that $\sigma (t)=e^{i\lambda (t)}I$
for $t\in T_0$ is
classical. Indeed, in the smooth model for $\pi$
treated in \S \ref{ssec:2.2}, the fact that the spectrum of $\pi$
is contained in the orbit of $\lambda$ implies that
$(\pi (t)f)(x)=e^{i\lambda (x^{-1}tx)}f(x)$ for all
$f\in {\bb}, t\in T_0, x\in G_0$. Hence $f(t^{-1})=
e^{i\lambda (t)}f(1)$ while $f(t^{-1})=\sigma (t)f(1)$.
So $\sigma (t)=e^{i\lambda (t)}I$.
\end{proof}
\begin{remark}
 In the classical theory {\it all\/} orbits of
$L_0$ are allowed and an additional argument of the positivity
of energy is needed to single out the physically occurring
representations. In SUSY theories as exemplified
by Theorem \ref{t3.1.1}, a restriction is already present:
only orbits in $T_0^+$ are permitted. We shall prove in
the next section that $T_0^+$ may be interpreted
{\it precisely\/} as the set of all
{\it positive energy representations.}
\end{remark}
\subsection{Determination of the admissible orbits.
Product structure of the representations
of the little super groups}
\label{ssec:3.2}
We fix
$\lambda \in T_0^+$ and let
$(\sigma ,\rho^\sigma )$ be a $\lambda$-admissible
irreducible UR of $S^\lambda$.
Clearly
$$
-id\sigma (Z)=\lambda (Z)I\qquad (Z\in \mathfrak t_0).
$$
Define
$$
\Phi_\lambda (X_1, X_2)= (1/2) \lambda ([X_1,X_2])\qquad
(X_1, X_2\in \mathfrak g_1).
$$
Then, on $C^\infty (\sigma )$,
$$
[\rho^\sigma (X_1),\rho^\sigma (X_2)]=\lambda ([X_1,X_2])I
=2 \Phi _\lambda (X_1, X_2)I.
$$
Clearly
$\Phi _\lambda $ is a symmetric bilinear form on $\mathfrak g_1
\times \mathfrak g_1$. Let
$$
Q_\lambda (X)= \Phi _\lambda(X, X)=(1/2)\lambda ([X,X]).
$$
Then $Q_\lambda $ is invariant
under $L^\lambda _0$ because for $X_1, X_2\in
\mathfrak g_1, h\in L_1$,
$$
[\rho^\sigma (hX_1),\rho^\sigma (hX_2)]=
\sigma (h)[\rho^\sigma (X_1),\rho^\sigma (X_2)]
\sigma (h)^{-1}= 2 \Phi _\lambda (X_1, X_2).
$$
Now
$$
\rho^\sigma (X)^2=Q_\lambda (X)I\qquad (X\in \mathfrak g_1).
$$
Since $\rho^\sigma (X)$ is essentially self adjoint
on $C^\infty (\sigma )$, it is immediate that
$Q_\lambda (X)\ge 0$. We thus obtain the necessary
condition for admissibility:
$$
Q_\lambda (X)=\Phi _\lambda (X, X)\ge 0
\qquad (X\in \mathfrak g_1).
$$
In the remainder of this subsection we shall
show that
the condition that $\Phi_\lambda\ge 0$,
which we refer to
as the {\it positive energy condition\/},
is also sufficient
to ensure that $\lambda $ is admissible. We will
then find all the $\lambda$-admissible
irreducible UR's of $S^\lambda $.

\medskip
It will follow in the next section that if
the super Lie group $(G_0, \mathfrak g_1)$ is a
super Poincar\'e group, the condition
$\Phi_\lambda \ge 0$ expresses precisely the positivity
of the energy. This is the reason for our describing this
condition in the general case also as the
positive energy condition.

\medskip
From now
on we fix $\lambda $ such that $\Phi_\lambda \ge 0$.
\begin{lemma}
\label{l3.2.1}
 For any admissible {\rm UR}
$(\sigma ,\rho^\sigma )$
of $S^\lambda $, $\rho^\sigma (X)$  is a bounded
self adjoint operator for $X\in \mathfrak g_1$,
and $\rho^\sigma (X)^2=
Q_\lambda (X)I$. Moreover, $Q_\lambda \ge 0$ and
is invariant under $L^\lambda _0$.
\end{lemma}
\begin{proof}
We have, for $X\in \mathfrak g_1, \psi
\in C^\infty (\sigma )$,
$$
|\rho^\sigma (X)\psi|_{\kk}^2=(\rho^\sigma (X)^2\psi ,\psi )
=Q_\lambda (X)^2|\psi |_{\kk}^2
$$
which proves the lemma.
\end{proof}

This lemma suggests we study the following
situation.
Let $W$ be a finite dimensional
{\it real\/} vector space and let $q$ be a
nonnegative quadratic form on $W$, i.e., $q(w)\ge 0$
for $w\in W$. Let $\varphi$ be the corresponding
symmetric bilinear form ($q(w)=\varphi (w,w)$).
Let ${\cc}$ be the real algebra generated by $W$ with the
relations $w^2=q(w)1 (w\in W)$. If $q$ is nondegenerate, i.e.,
positive definite, this is the
Clifford algebra associated to the quadratic vector space
$(W, q)$. If $q=0$ it is just the exterior algebra over
$W$. If $(w_i)_{1\le i\le n}$ is a basis for $W$ such
that $\varphi (w_i, w_j)=\varepsilon _i\delta _{ij}$ with
$\varepsilon _i=0$ or $1$ according as $i\le a$ or $>a$,
then ${\cc}$ is the algebra generated by the $w_i$ with
the relations $w_iw_j+w_jw_i=2\varepsilon _i\delta _{ij}$.
Let $W_0$ be the radical of $q$, i.e.,
$W_0=\{w_0|\varphi (w_0, w)=0\hbox { for all }w\in W\}$;
in the above notation $W_0$ is spanned by the
$w_i$ for $i\le a$. If $W^\sim=W/W_0$
and $q^\sim , \varphi ^\sim$
are the corresponding objects induced on
$W^\sim$, $q^\sim$
is positive definite, and so we have the
usual Clifford algebra
${\cc}^\sim $ generated by $(W^\sim , q^\sim )$
with $W^\sim \subset {\cc}^\sim$. The
natural map $W\longrightarrow
W^\sim$ extends uniquely to a morphism
${\cc}\longrightarrow
{\cc}^\sim$ which is clearly surjective.
We claim that its kernel is the ideal
${\cc}_0$ in ${\cc}$ generated by $W_0$. Indeed, let $I$
be this kernel. If $s\in I$, $s$ is a linear
combination of elements $w_Iw_J$ where $w_I$
is a product $w_{i_1}\dots w_{i_r} (i_1<\dots <i_r\le a)$
and $w_J$ is a product
$w_{j_1}\dots w_{j_s} (a<j_1<\dots <j_s)$; hence,
$s\equiv \sum _Jc_Jw_J$ mod ${\cc}_0$, and as the
image of this element in ${\cc}^\sim $ is $0$,
$c_J=0$ for all $J$ because the images
of the $w_J$ are linearly independent in ${\cc}^\sim$.
Hence $s\in {\cc}_0$, proving our claim.

\medskip
A representation $\theta$ of ${\cc}$ by
{\it bounded\/} operators in
a SHS ${\kk}$ is called {\it self adjoint\/} (SA) if
$\theta (w)$ is odd and self adjoint for all $w\in W$.
$\theta$ can be viewed as a representation of the
complexification ${\complex}\otimes {\cc}$ of ${\cc}$; a
representation of ${\complex}\otimes {\cc}$ arises in this
manner from a SA representation of ${\cc}$ if and only if
it maps elements of $W$ into odd operators and
takes complex conjugates to adjoints. Also we wish to stress
that irreducibility is in the graded sense.
\begin{lemma}
\label{l3.2.2}
 {\rm (}i{\rm )} If $\tau$ is a
{\rm SA} representation of ${\cc}$ in ${\kk}$, then $\tau =0$
on ${\cc}_0$ and so it is the lift of a {\rm SA} representation
$\tau ^\sim $ of ${\cc}^\sim$. {\rm (}ii{\rm )}
There exist irreducible
{\rm SA} representations $\tau$ of ${\cc}$; these are finite
dimensional, unique if $\dim (W^\sim)$ is odd, and unique up
to parity reversal if $\dim (W^\sim )$ is even.
{\rm (}iii{\rm )} Let $\tau$ be an irreducible {\rm SA}
representation of ${\cc}$ in a {\rm SHS} ${\ll}$
and let $\theta $
be any {\rm SA} representation of ${\cc}$ in a
{\rm SHS} ${\rr}$. Then
${\rr}\simeq {\kk}\otimes {\ll}$ where ${\kk}$ is a {\rm SHS}
and $\theta (a)=1\otimes \tau (a)$ for all $a\in {\cc}$; moreover,
if $\dim (W^\sim )$ is odd, we can choose ${\kk}$ to
be purely even.
\end{lemma}
\begin{proof}
 (i) If $w\in W_0$, then $\tau (w_0)^2=q(w_0)I=0$
and so, $\tau (w_0)$ itself must be $0$ since it is self adjoint.

\medskip
(ii) In view of (i) we may assume that $W_0=0$ so that $q$
is positive definite.

\medskip\noindent
{\it Case I {\rm :} $\dim (W)=2m$.\/} Select an ON basis
$a_1, b_1, \dots , a_m, b_m$ for $W$. Let $e_j=(1/2)(a_j+ib_j),
f_j=(1/2)(a_j-ib_j)$. Then $ \varphi(e_j, e_k)=
\varphi(f_j, f_k)=0$ while $ \varphi(e_j, f_k)=(1/2)\delta _{jk}$.
Then ${\complex}\otimes {\cc}$ is generated by the $e_j, f_k$
with the relations
$$
e_je_k+e_ke_j=f_jf_k+f_kf_j=0,\quad  e_jf_k+f_ke_j=\delta _{jk}.
$$
We now set up the standard \lq\lq Schr\"odinger\rq\rq\
representation of ${\complex}\otimes {\cc}$. The representation
acts on the SHS ${\ll}=\Lambda (U)$ where $U$ is a
Hilbert
space of dimension $m$ and the grading on ${\ll}$ is
the ${\z}_2$-grading induced by the usual
${\z}$-grading of ${\Lambda(U)}$. Let $(u_j)_{1\le j\le m}$
be an ON
basis for $U$. We define
$$
\tau (e_j)f=u_j\wedge f, \qquad \tau (f_j)f=\partial (u_j)(f)
\qquad (f\in \Lambda (U)),
$$
$\partial (u)$ for any $u\in U$ being the odd derivation on
$\Lambda (U)$ such that $\partial (u)v= 2 (v,u)$ (here $\left(\cdot,\cdot\right)$ is the scalar product in $\Lambda(U)$ extending the scalar product of $U$). It is standard
that $\tau$ is an irreducible representation of
${\complex}\otimes {\cc}$. The vector $1$ is called
the {\it Clifford vacuum.\/} We shall now verify
that $\tau$ is SA
Since $a_j=e_j+f_j, b_j=-i(e_j-f_j)$, we need to verify
that $\tau (f_j)=\tau (e_j)^\ast$ for all $j$, $\ast$ denoting
adjoints. For any subset $K=\{k_1<\dots <k_r\}\subset
\{1,2,\dots ,m\}$ we write $u_K=u_{k_1}\wedge \dots \wedge
u_{k_r}$. Then we should verify that
$$
(u_j\wedge u_K, u_L)=(u_K, \partial (u_j)u_L)
\qquad (K, L\subset \{1,2,\dots ,m\}).
$$
Write $K=\{k_1,\dots ,k_r\}, L=\{\ell_1,\dots ,\ell_s\}$
where $k_1<\dots <k_r, \ell_1<\dots <\ell_s$. We assume
that $j=\ell_a$ for some $a$ and $K=L\setminus \{\ell_a\}$,
as otherwise both sides are $0$. Then $K=\{\ell_1,\dots ,
\ell_{a-1},\ell_{a+1},\dots ,\ell_s\}$ (note that
$r=s-1$). But then both sides are equal to $(-1)^{a-1}$.

\medskip
From the general theory of Clifford algebras we know that
if $\tau'$ is another
irreducible SA representation of ${\cc}$,
then either $\tau \approx \tau'$ or else $\tau \approx
\Pi\tau'\Pi$ where $\Pi$ is the parity revarsal map and we write $\approx$ for linear (not
necessarily unitary) equivalence. So it remains to
show that $\approx$ implies unitary equivalence which
we write $\simeq$. This is standard since
the linear equivalence preserves self adjointness.
Indeed, if $R:\tau_1\longrightarrow
\tau_2$ is an even linear isomorphism, then $R^\ast R$ is
an even automorphism of $\tau_1$ and so $R^\ast R=a^2I$
where $a$ is a scalar which is $>0$. Then $U=a^{-1}R$
is an even unitary isomorphism
$\tau_1\simeq \tau_2$.  Also for
use in the odd case to be treated next, we note that $\tau$
is irreducible in the ungraded sense since its
image is the full endomorphism algebra of ${\ll}$.

\medskip
{\it Case II {\rm :} $\dim (W)=2m+1$.\/} It is enough
to construct an irreducible SA representation as it
will be unique up to linear, and hence unitary,
equivalence.

\medskip
Let $a_0,a_1,\dots ,a_{2m}$ be an ON basis for $W$.
Write $x_j=ia_0a_j (1\le j\le 2m), \quad x_0=i^ma_0a_1
\dots  a_{2m}$. Then $x_0^2=1, x_jx_k+x_kx_j=2
\delta _{jk} (
j,k=1,2,\dots ,2m)$. Moreover $x_0$ commutes with all $a_j$
and hence with all $x_j$. The $x_j$ generate a Clifford
algebra over ${\real}$ corresponding to a positive definite
quadratic form and so there is an irreducible ungraded
representation $\tau^+$ of it in an ungraded Hilbert space ${\ll}^+$
such that $\tau^+(x_j)$ is self adjoint for all
$j=1,2,\dots ,2m$ (cf. remark above). Within ${\complex}\otimes {\cc}$ the $x_j$
generate ${\complex}\otimes {\cc}^+$ so that $\tau^+$ is a
representation of ${\complex}\otimes {\cc}^+$ in ${\ll}^+$ such
that $i\tau^+(a_0a_j)$ is self adjoint for all $j$.
We now take
$$
{\ll}={\ll}^+\oplus {\ll}^+,\quad \tau=
\tau^+\oplus \tau^+, \quad \tau(x_0)=
\begin{pmatrix}
0&1\cr 1&0\cr
\end{pmatrix}.
$$
Here ${\ll}$ is given the ${\z}_2$-grading
such that the first and second copies of ${\ll}^+$
are the even and odd parts.
It is clear that $\tau$ is an irreducible representation
of ${\complex}\otimes {\cc}$. We wish to show that
$\tau(a_r)$ is odd and self adjoint for
$0\le r\le 2m$. But this follows from the fact that
the $\tau (x_r)$ are self adjoint, $\tau (x_j)$ are even,
and $\tau (x_0)$ is odd, in view of the formulae
$$
a_0=i^mx_0x_1\dots x_{2m}, \quad a_j=-ia_0x_j.
$$
This finishes the proof of (ii).

\medskip
(iii) Let now $\theta $ be a SA representation of ${\cc}$
in a SHS ${\rr}$ of {\it possibly infinite\/}
dimension. For any homogeneous
$\psi \in {\rr}$ the cyclic subspace $\theta ({\cc})\psi$
is finite dimensional, hence closed, graded and is
$\theta$-stable; moreover by the SA nature of $\theta$,
for any graded invariant subspace its orthogonal complement
is also graded and invariant. Hence we can write
${\rr}=\oplus _\alpha
{\rr}_\alpha$ where the sum is direct and each ${\rr}_\alpha$
is graded, invariant, and irreducible. Let ${\ll}$ be
a SHS on which we have an irreducible SA representation
of ${\cc}$.
If $\dim (W)^\sim$ is even we
can thus write ${\rr}=({\mm}_0\otimes {\ll})\oplus
({\mm}_1\otimes \Pi {\ll})$ where the ${\mm}_j$
are even Hilbert spaces; if $\dim (W^\sim)$
is odd we can write ${\rr}={\kk}\otimes {\ll}$ where
${\kk}$ is an even Hilbert space. In the first case,
since ${\mm}_1\otimes \Pi {\ll} =
\Pi {\mm}_1\otimes {\ll}$, we
have ${\rr}={\kk}\otimes {\ll}$ where ${\kk}$
is a SHS with ${\kk}_0={\mm}_0, {\kk}_1=\Pi {\mm}_1$.
\end{proof}

For studying the question of admissibility
of $\lambda$ we need a second ingredient.
Let $H$ be a {\it not necessarily connected\/}
Lie group and let us
be given a morphism
$$
j : H\longrightarrow {\rm O}
(W^\sim )
$$
so that $H$ acts on $W^\sim$
preserving the quadratic form on $W^\sim$.
We wish to find out when there is a UR $\kappa$
of $H$, {\it possibly
projective, and preferably, but not necessarily, even\/},
in the space of the irreducible SA representation
$\tau^\sim $, such that
$$
\kappa (t)\tau^\sim (w)\kappa (t)^{-1}=
\tau^\sim (tw)\qquad (t\in H, w\in W^\sim )
\eqno (\ast ).
$$
For $h\in {\rm O}(W^\sim)$, let
$$
\tau ^\sim _h(w)=\tau ^\sim (hw)\qquad (w\in W^\sim).
$$
Then $\tau^\sim_h$ is also an irreducible SA
representation of ${\cc}^\sim$ and so we can find a
unitary operator $K(h)$ such that
$$
\tau^\sim_h(w)=K(h)\tau^\sim (w)K(h)^{-1}\qquad (w\in W^\sim).
$$
If $\dim W^\sim$ is even, $\tau^\sim$ is irreducible
even as an ungraded representation, and so
$K(h)$ will be unique up to a phase;
it will be even or odd
according as $\tau^\sim_h\simeq \tau^\sim$ or
$\tau^\sim_h\simeq \Pi\tau^\sim\Pi$ where $\Pi$
is parity reversal. If $\dim W^\sim$ is odd,
$\tau^\sim$ is irreducible only as a graded
representation and so we
also need to require $K(h)$ to be an {\it even} operator
in order that it is uniquely determined up to a phase.
With this additional requirement in the odd dimensional case,
we then see that in both cases the class of $\kappa$ as a
projective UR of $H$ is uniquely determined, i.e., the class
of its multiplier $\mu$ in $H^2(H,\tor)$ is fixed. In the following,
we shall show that $\mu$ can be chosen to
be $\pm 1$-valued, and
examine the structure of $\kappa$ more closely.

 \medskip
We begin with some preparation (see \cite{D},\cite{V}).
Let ${\cal C}^{{\sim \times }}$ be the group of
invertible elements in ${\cal C}^{{\sim }}$.
Define the {\it full Clifford group} as follows:
$$
\Gamma =\left\{ x\in {\cal C}^{{\sim \times }}\cap ({\cal C}^{{\sim +}}
\cup {\cal C}^{{\sim -}})\mid x W^{\sim } x^{-1}\subset W^{\sim }\right\}.
$$
We have a homomorphism $\alpha :\Gamma \longrightarrow
\rm{O}(W^{\sim })$ given by
$$
\alpha (x)w =(-1)^{p(x)}xwx^{-1}
$$
for all $w\in W^{\sim }$, $p(x)$ being $0$ or $1$
according as $x\in {{\cc}^\sim}^+$ or $x\in {{\cc}^\sim}^-$.
Let $\beta $ be the
principal antiautomorphism of ${\cal C}^{{\sim +}}$;
then $x\beta (x)\in {\real}^\times$ for all $x\in \Gamma$, and
we write $G$ for the kernel of the homomorphism
$x\mapsto x\beta (x)$ of $\Gamma$ into
${\real}^\times$. Since $W^{\sim }$ is a positive
definite quadratic space, we
have an exact sequence
$$
1\longrightarrow \{\pm 1\}\longrightarrow G
\smash{\mathop{\longrightarrow}
\limits^{\alpha}}  {\rm O}(W^{\sim
})\longrightarrow 1.
$$
For $\dim W^{\sim }\geq 2$, the connected component $G^{0}$ of $
G$ is contained in ${\cal C}^{{\sim +}}$ and coincides with
${\rm Spin}(W^{\sim })$.
\begin{lemma}
\label{l3.2.3}
Let $\tau ^{\sim }$ be a SA irreducible representation
of ${\cal C}^{{\sim }}$. We then have the following.
{\rm {(}i\rm {)} }$\tau ^{\sim }$ restricts to a
unitary representation of $G$. {\rm {(}ii\rm {)}} The
operator $\tau ^{\sim }\left( x\right) $ is even or odd according
as $x\in G^{0}$ or $x\in G\setminus G^{0}$. {\rm {(}iii{)}}
$\tau^\sim (x)\tau^\sim (w)\tau^\sim (x)^{-1}
=(-1)^{p(x)}\tau^\sim (\alpha (x)(w))$
for $x\in G, w\in W^\sim$.
\end{lemma}
\begin{proof}
Each $x\in G$ is expressible in the form $x=cv_{1}\ldots v_{r}$, where
$v_{i}$ are unit vectors in $W^{\sim }$ and $c\in \{\pm 1\}$.
Since $\tau ^{\sim
}\left( v_{i}\right) $ is odd, the parity of $\tau ^{\sim }\left(
x\right)$
is the same of $x$. Moreover, since the $\tau ^{\sim }\left( v_{i}\right)$
are self adjoint,
$$
\tau ^{\sim }\left( x\right) \tau ^{\sim }
\left( x\right)^{\ast }
=c^{2}\tau
^{\sim }\left( v_{1}\right) \ldots \tau ^{\sim }
\left( v_{r}\right) \tau
^{\sim }\left( v_{r}\right)\ldots \tau ^{\sim }
\left( v_{1}\right)
=I .
$$
This proves (i). (ii) and (iii) are obvious.
\end{proof}
We now consider two cases.

{\it Case {\rm I\/}.\/} $j(H)\subset {\rm SO}(W^\sim )$.\\
Let $\zeta $ be a Borel
map of ${\rm SO}(W^\sim )$ into ${\rm Spin}(W^\sim )$
which is a right inverse of $\alpha ({\rm Spin}(W^\sim )
\longrightarrow {\rm SO}(W^\sim ))$ with $\zeta (1)=1$.
Then $\zeta (xy)=\pm \zeta (x)\zeta (y)$
for $x, y \in {\rm SO}(W^\sim )$, and so
$$
\kappa_H=\tau^\sim \circ \zeta \circ j
$$
is an even projective UR of $H$ satisfying $(\ast )$ with
a $\pm$-valued multiplier $\mu_H$.
Since $\zeta (1)=1$ it follows that $\mu_H$ is
{\it normalized\/}, i.e.,
$$
\mu_H(h,1)=\mu_H(1,h)=1\qquad (h\in H).
$$
Clearly, the class of $\mu_H$ in $H^2(H, {\z}_2)$
is trivial if and only if $j: H\rightarrow {\rm SO}(W^\sim)$
can be lifted to a morphism $\widehat{j}: H\rightarrow
{\rm Spin}(W^\sim)$.
In particular, this happens if $H$ is connected and
simply connected.

\medskip
Suppose now $H$ is connected but $\widehat j$
does not exist. We
can then find a two-fold cover $H^\sim$ of $H$ with a
covering map $p (H^\sim \longrightarrow H)$ such that
$j(H\rightarrow {\rm SO}(W^\sim))$ lifts to a morphism
$j^\sim (H^\sim \rightarrow {\rm Spin}(W^\sim))$, and
if $\xi$ is the nontrivial element in
$\ker p$, then $j^\sim (\xi) = -1$.
\begin{lemma}
\label{l3.2.4}
If $j$ maps $H$
into ${\rm SO}(W^\sim )$, there is a projectively unique even
projective {\rm UR} $\kappa$ of $H$ satisfying
$(\ast )$, with a normalized $\pm 1$-valued
multiplier $\mu$. If $H$ is connected, for
$\kappa$ to be an ordinary even representation
{\rm (}which will be unique up to multiplication by a
character of $H${\rm )}
it is necessary and sufficient that either
{\rm (}i{\rm )} $j(H\rightarrow {\rm SO}(W^\sim))$
can be lifted to ${\rm Spin}(W^\sim)$ or {\rm (}ii{\rm )}
there exists a
character $\chi$ of $H^\sim$ such that
$\chi (\xi )=-1$. In particular,
if $H=A\times'T$, where $A$ is simply connected and
$T$ is a torus, then $\kappa $ is an ordinary even
unitary representation.
\end{lemma}
\begin{proof}
 The first statement has already been proved.

\medskip
We next prove the sufficiency part of the second statement.
Sufficiency of {\rm (}i{\rm )} has already been observed.
To see that (ii) is sufficient, note that $\kappa^\sim =
\tau^\sim \circ j^\sim$ is an even UR of $H^\sim$
satisfying $(\ast )$; one can clearly replace
$\kappa^\sim$ by $\kappa^\sim \chi$ without
destroying $(\ast )$. As $\kappa^\sim (\xi)=-1$, we
have $(\kappa^\sim \chi)(\xi)=1$, and so it is immediate
that $\kappa^\sim \chi$ descends to $H$.

\medskip
We leave the necessity part to the reader;
it will not be used in the sequel.

\medskip
The statement for $H=A\times'T$ will follow if we show that $H^\sim$ has a
character $\chi$ as in condition {\rm (}ii{\rm )}. We have
$H^\sim =A\times'T^\sim$, $T^\sim$ being the double cover of $T$, and
$\xi = (1,t)$, with $t\not=1, t^2 =1$. There exists a
character $\chi$ of $T^\sim$
such that $\chi (t) = -1$, and such a character can be extended to
$H^\sim$ by
making it trivial on $A$.
\end{proof}

\medskip
{\it Case {\rm II\/}.\/} $j(H)\not\subset
{\rm SO}(W^\sim )$.\\
Let
$$
H_0=j^{-1}({\rm SO}(W^\sim )).
$$
Then $H_0$ is a normal subgroup of $H$ of index $2$.
We must distinguish two subcases.

\medskip
{\it Case {\rm II.a\/}.\/} $\dim (W^\sim)$ is even.\\
Let $\zeta_0$ be a Borel right inverse of
$\alpha (G^0\longrightarrow {\rm SO}(W^\sim ))$
with $\zeta_0 (1) = 1$.
Fix a unit vector $v_0\in W^\sim$ and let $r_0 = - \alpha (v_0)$.
Since $\alpha (v_0)$ is the reflection in the hyperplane
orthogonal to $v_0$ and $\dim (W^\sim)$ is even, we see that
$r_0 \in {\rm O}(W^\sim) \setminus {\rm SO}(W^\sim)$. We then
define a map $\zeta ({\rm O}(W^\sim) \rightarrow G)$ by
$$
\zeta (h)=
\begin{cases}
\zeta _0(h)& \mbox{ if } h\in {\rm SO}(W^\sim )\cr
\zeta_0(h_0)v_0& \mbox{ if } h=h_0r_0, h_0\in {\rm
SO}(W^\sim ) .\cr
\end{cases}
$$
Once again we have $\zeta (h_1h_2)=\pm \zeta (h_1)\zeta (h_2)$.
Define
$$
\kappa_H=\tau^\sim \circ \zeta \circ j.
$$
Then $\kappa _H$ is a projective UR of $H$ satisfying $(\ast )$
with a $\pm$-valued normalized multiplier $\mu_H$.
But $\kappa_H$ is {\it not\/}
an even representation; elements of $H\setminus H_0$ map into
{\it odd\/} unitary operators in the space of $\tau^\sim$. We
shall call
such a representation of $H$ {\it graded with respect to $H_0$\/},
or simply {\it graded.\/} We have thus proved the following.
\begin{lemma}
\label{l3.2.5}
If $j(H)\not\subset {\rm SO}(W^\sim )$,
and $\dim (W^\sim )$ is even, and if
we define $H_0=j^{-1}({\rm SO}(W^\sim ))$, then there is
a projective {\rm UR} $\kappa_H$, graded with respect
to $H_0$, and satisfying $(\ast )$ with
a $\pm$-valued normalized multiplier $\mu_H$.
\end{lemma}

Before we take up the case when $\dim (W^\sim)$ is odd, we shall
describe how the projective graded representations of $H$
are constructed. This is a very general situation and so we
shall work with a locally compact second countable group $A$
and a closed subgroup $A_0$ of index $2$;
$A_0$ is automatically
normal and we write $A_1=A\setminus A_0$. Gradedness
is with respect to $A_0$. We fix a $\pm$-valued
multiplier $\mu$ for $A$ which is normalized.
For brevity a representation will mean a unitary $\mu$-representation.
Moreover, with a slight abuse of language a $\mu$-representation of $A_0$
will mean a $\left. \mu \right|_{A_0 \times A_0}$-representation of $A_0$.
If $R_g$ is a graded representation
in a SHS ${\hh}$, $R$ the corresponding ungraded
representation, and $P_j$ is the orthogonal projection
${\hh}\longrightarrow {\hh}_j$, we associate to
$R$ the projection valued measure $P$
on $A/A_0$ where $P_{A_0}=P_0$ and $P_{A_1}=P_1$. Then
the condition that $R_g$ is graded is exactly the same
as saying that $(R,P)$ is a system of imprimitivity
for $A$ based on $A/A_0$. Conversely, given a system
of imprimitivity $(R, P)$ for $A$ based on $A/A_0$,
let us define the grading for ${\hh}={\hh}(R)$
by ${\hh}_j=\hbox { range of }P_{A_j} (j=0,1)$; then
$R$ becomes a graded representation. Moreover
for graded representations $R_g, R'_g$, we have
${\rm Hom}(R_g,R'_g)={\rm Hom}((R,P),(R',P'))$.
In other words, the category of systems of
imprimitivity for $A$ based on $A/A_0$ and the category
of representations of $A$ graded with respect to
$A_0$ are equivalent naturally.

\medskip
For any $\mu$-representation $r$
of $A_0$ in a purely even Hilbert space ${\hh}(r)$, let
$$
R_r:={\rm Ind}_{A_0}^A\,r
$$
be the representation of $A$ induced by $r$. We recall that
$R_r$ acts in the Hilbert space ${\hh}(R_r)$ of all
(equivalence classes of Borel) functions $f(A\rightarrow
{\hh}(r))$ such that for each $\alpha \in A_0$,
$$
f(\alpha a)=\mu(\alpha ,a)r(\alpha )f(a)
$$
for almost all $a\in A$; and
$$
(R_r(a)f)(y)=\mu (y,a)f(ya)\qquad (a, y\in A).
$$
The
space ${\hh}(R_r)$ is naturally graded by defining
$$
{\hh}(R_r)_j=\{f\in {\hh}(R_r)\ |\ {\rm supp}
(f)\subset A_j\}\qquad (j=0,1).
$$
It is then obvious that $R_r$ is a graded
$\mu$-representation. We write
$\widehat {R_r}$ for $R_r$ treated as a graded
representation. These remarks suggest the following lemma.
\begin{lemma}
\label{l3.2.6}
For any unitary
$\mu$-representation
$r$ of $A_0$ let $R_r={\rm Ind\/}(r)$
and let $\widehat {R_r}$ be the graded
$\mu$-representation defined by
$R_r$. Then the assignment
$r\mapsto \widehat {R_r}$ is an equivalence from the
category of unitary $\mu$-representations of $A_0$
to the category of unitary graded $\mu$-representations
of $A$.
\end{lemma}
\begin{proof}
 Let us first assume that $\mu=1$. Then
we are dealing with UR's and the above remarks imply
the Lemma in view of the classical imprimitivity theorem.

When $\mu$ is not $1$ we go to the central extension
$A^\sim$ of $A$ by ${\z}_2$ defined by $\mu$. Recall
that $A^\sim =A\times _\mu{\z}_2$ with multiplication defined by
$$
(a,\xi )(a',\xi')=(aa', \xi \xi'\mu (a,a'))\qquad
(a,a'\in A, \xi ,\xi'\in {\z}_2).
$$
(We must give to $A^\sim$ the Weil topology).
Then $A_0^\sim =A_0\times _\mu{\z}_2$ and
$A^\sim /A_0^\sim =A/A_0$.
The  $\mu$-representations $R$ of $A$ are in natural
bijection with UR's $R^\sim$ of $A^\sim$ such that $R^\sim $ is
nontrivial on $\z_2$ by the
correspondence
$$
R^\sim (a,\xi )=\xi R(a), \quad R(a)=R^\sim (a,1).
$$
The assignment $R\mapsto R^\sim$ is an equivalence
of categories. Analogous considerations hold for $\mu$-representations
$r$ of $A_0$ and UR's $r^\sim$ of $A_0^\sim$ which are
nontrivial on $\z_2$. The Lemma would now follow if we
establish two things: (a) For any unitary
$\mu$-representation $r$ of $A_0$, and $R_r=
{\rm Ind\/}(r)$, we have
$$
R_r^\sim \simeq {\rm Ind\/}(r^\sim )
$$
and (b) If $\rho$ is a UR of $A_0^\sim $
and ${\rm Ind\/}(\rho)=R^\sim$ for some
$\mu$-representation $R$ of $A$, then
$\rho = r^\sim $ for some $\mu$-representation
$r$ of $A_0$. To prove (a) we set up the map
$f\longmapsto f^\sim$ from ${\hh}(R_r^\sim )$ to
${\hh}({\rm Ind\/}(r^\sim ))$ by
$$
f^\sim (a,\xi )=f(a)\xi.
$$
It is an easy calculation that this is
an isomorphism of $R_r^\sim$ with
${\hh}({\rm Ind\/}(r^\sim ))$ that intertwines the two
projection valued measures on $A/A_0$ and
$A^\sim /A_0^\sim\approx A/A_0$. To prove (b)
we have only to check that $\rho (1,\xi )=\xi$; this
however is a straightforward calculation.
\end{proof}
\begin{remark}
 Given a graded $\mu$-representation
$R$ of $A$, let $r$ be the
$\mu$-representation of $A_0$ defined by
$$
r(\alpha )=R(\alpha )\big|_{{\hh}(R)_0}\qquad (\alpha \in A_0).
$$
It is then easy to show that $R\simeq R_r$.
In fact it is enough to verify this (as before) when
$\mu=1$. In this simple situation this is well known.
\end{remark}
We now resume our discussion and treat the
odd dimensional case.

{\it Case {\rm II.b\/}.\/} $\dim (W^\sim)$ is odd.\\
We shall exhibit a projective even UR $\kappa$ of $H$
satisfying $(\ast)$. We refer back to the construction
of $\tau^\sim$ in Lemma \ref{l3.2.2}. Then
$$
S=\begin{pmatrix}
0&1\cr -1&0\cr
\end{pmatrix}
$$
is an odd unitary operator such that
$S^2=-1$ and $\tau^\sim (x) S = (-1)^{p(x)} S \tau^\sim (x)$ for all $x\in
\cc^\sim$.
Let $\gamma ({\rm O}(W^\sim) \rightarrow G)$ be a Borel right inverse of
$\alpha$.
Then, it is easily checked that
$$
\kappa_H \left( h\right) =
\begin{cases}
\left(\tau^\sim \circ \gamma \circ j \right) \left( h\right) & \mbox{ if }
h\in H_{0}\cr
\left(\tau^\sim \circ \gamma \circ j \right) \left( h\right) S & \mbox{ if }  h\in
H\setminus H_{0}\cr
\end{cases}
$$
has the required properties.

\bigskip
We now return to our original setting. First
the graded representations of $H$ are obtained by
taking $H=A, H_0=A_0$ in the foregoing discussion.
Let $\mathfrak g_{1\lambda }=\mathfrak g_1/{\rm rad}
\ \Phi_\lambda $. We write
${\cc}_\lambda $ for the algebra generated by $\mathfrak g_1$
with the relations $X^2=Q_\lambda (X)1$ for
all $X\in \mathfrak g_1$. We have a map
$$
j_\lambda : L^\lambda _0\longrightarrow
{\rm O}(\mathfrak g_{1\lambda }).
$$
We take in the preceding theory
$$
q=Q_\lambda, \quad \varphi = \Phi_\lambda, \quad H=L^\lambda _0, \quad W^\sim =\mathfrak g_{1\lambda },
\quad j=j_\lambda ,\quad {\cc}^\sim ={\cc}^\sim _\lambda.
$$
Furthermore let
$$
\kappa _\lambda =\kappa,\quad \mu_\lambda =\mu,
\quad \tau_\lambda^\sim =\tau^\sim, \quad \tau_\lambda =
\hbox { lift of }\tau^\sim \hbox { to } {\cc}_\lambda .
$$
Then $\mu_\lambda$ is a normalized
multiplier for $L^\lambda_0$ which we can choose to be
$\pm$-valued, $\kappa_\lambda$
is a $\mu_\lambda$-representation (unitary) of
$L^\lambda_0$ in the space of $\tau_\lambda$, and
$$
\kappa_\lambda (t)\tau _\lambda (X)
\kappa_\lambda (t)^{-1}=\tau_\lambda (tX)\qquad
(t\in L^\lambda _0).
$$
Moreover, $\kappa_\lambda$ is graded if and only if
$j_\lambda (L^\lambda _0)\not\subset {\rm SO}
(\mathfrak g_{1\lambda })$ and $\dim (\mathfrak g_{1\lambda })$ is even,
otherwise $\kappa_\lambda$ is even. Finally, let
$$
L^\lambda_{00}=\begin{cases}
j_\lambda ^{-1}({\rm SO}
(\mathfrak g_{1\lambda })) & \mbox{ if }j_\lambda (L^\lambda _0)\not\subset {\rm SO}
(\mathfrak g_{1\lambda }) \mbox{ and }\dim (\mathfrak g_{1\lambda }) \mbox{ is even } \cr
L^\lambda _0 & \mbox{ otherwise}.\cr
\end{cases}
$$
For any
unitary $\mu_\lambda$-representation $r$ of
$L^\lambda _{00}$ in an even Hilbert
space ${\kk}_\lambda $, let
$\widehat {R_r}$ be the unitary $\mu_\lambda$-
representation of $L^\lambda_0$ induced by $r$, which
is graded if $j_\lambda (L^\lambda _0)\not\subset {\rm SO}
(\mathfrak g_{1\lambda })$ and $\dim (\mathfrak g_{1\lambda })$ is even,
and is just $r$ in all other cases.
\begin{theorem}
\label{t3.2.7}
Let $\lambda $ be such that
$\Phi_\lambda \ge 0$. Then $\lambda $ is
admissible, i.e.,
$\lambda \in T_0^+$.
For a fixed such $\lambda $ let $\tau_\lambda $ be an
irreducible {\rm SA} representation of ${\cc}_\lambda $ in
a {\rm SHS} ${\ll}_\lambda $ and $\kappa_\lambda $ the
unitary $\mu_\lambda$-representation of $L^\lambda _0$
in ${\ll}_\lambda $ associated to $\tau_\lambda$ as above. For any
unitary $\mu_\lambda$-representation $r$ of
$L^\lambda _{00}$ in an even Hilbert
space ${\kk}_\lambda $, let
$\widehat {R_r}$ be the unitary $\mu_\lambda$-
representation of $L^\lambda_0$ defined as above, and let
$$
\theta _{r\lambda }=
(\sigma_{r\lambda },
\rho_\lambda ^\sigma)
$$
be the {\rm UR} of the little {\rm
SLG} $S^\lambda $, where,
for $X\in \mathfrak g_1, h\in L^\lambda _0, t\in T_0$,
$$
\sigma _{r\lambda }(th)=e^{i\lambda (t)}\sigma '
_{r\lambda }(h)
$$
and
$$
\sigma '_{r\lambda }(h)=
\widehat {R_r}(h)\otimes \kappa_\lambda (h),
\quad \rho_\lambda ^\sigma (X)=1\otimes
\tau_\lambda (X)\qquad (X\in \mathfrak
g_1).
$$
Then $\theta _{r\lambda }$ is an admissible
{\rm UR} of $S^\lambda $. The assignment
$r\longmapsto \theta _{r\lambda }$
is functorial, commutes with direct sums,
and is an equivalence of categories
from the category of unitary $\mu_\lambda$-representations
of $L^\lambda _{00}$
to the category of admissible {\rm UR}'s
of the little super group $S^\lambda$.
If $L^\lambda _0$ is connected and satisfies either of
the conditions of Lemma \ref{l3.2.4}, then $r\longmapsto \theta _{r\lambda }$
is an equivalence from the category of even {\rm UR}'s of $L^\lambda _0$
into the category of admissible {\rm UR}'s of $S^\lambda$.
\end{theorem}
\begin{proof}
 Once $\kappa _\lambda $ is fixed, the
assignment
$$
r\longmapsto \theta_{r\lambda}
$$
is clearly functorial (although it
depends on $\kappa_\lambda $).
If $\dim(\mathfrak g_{1\lambda })$ is even, a morphism
$M:r_1\longrightarrow r_2$
obviously gives rise
to the morphism $\widehat M: {R_{r_1}}
\longrightarrow {R_{r_2}}$ and hence to the morphism
$\widehat M\otimes 1$ from $\theta_{r_1 \lambda}$
to $\theta_{r_2 \lambda}$. Conversely, if $T$
is a bounded even operator commuting with $1\otimes \tau_\lambda$, it is immediate (since the
$\tau_\lambda (X)$ generate the full super algebra
of endomorphisms of ${\ll}_\lambda$) that
$T$ must be of the form $M'\otimes 1$ where
$M':{\hh} (\widehat {R}_{r_1})\rightarrow
{\hh} (\widehat {R}_{r_2})$
is a bounded even operator. If now $T$ intertwines
$\widehat {R}_{r_1}\otimes \kappa_\lambda$ and $\widehat
{R}_{r_2}\otimes
\kappa_\lambda$, then $M'$ must belong to
${\rm Hom}(\widehat {R}_{r_1}, \widehat {R}_{r_2})\approx {\rm Hom} (r_1,
r_2)$. Thus $r\longmapsto
\theta _{r\lambda }$ is a fully faithful
functor. If $\dim(\mathfrak g_{1\lambda })$ is odd, we can
choose ${\kk}_\lambda$ to be purely even (see Lemma \ref{l3.2.2}).
If $T$ is a bounded even operator commuting
with $1\otimes \tau _\lambda$, we use the fact that it
commutes with
$$
1\otimes \begin{pmatrix}
\tau^+(a)&0\cr 0&\tau^+(a)\cr
\end{pmatrix}
\hbox { and }
1\otimes \begin{pmatrix} 0&1\cr 1&0\cr
\end{pmatrix}
$$
(in the notation of Lemma \ref{l3.2.2}) to conclude, via an
argument similar to the one used in the even dimensional case,
that $T$ is of the form $M\otimes 1$. Arguing as before we conclude
that $M\in {\rm Hom}(r_1, r_2)$. Thus
$r\mapsto\theta_{r\lambda}$ is a fully faithful functor
in this case also.
It remains to
show that every admissible UR of $S^\lambda $ is of the
form $\theta _{r\lambda }$.

\medskip
Let $\theta$ be an admissible UR of $S^\lambda $ in ${\hh}$.
Then $\theta =(\xi ,\tau)$ where $\xi$ is an even UR of
$T_0L^\lambda _0$ which restricts to $e^{i\lambda }I$ on $T_0$,
$\tau$ is a SA representation of ${\cc}_\lambda $ related
to $\xi$ as usual. We may then assume
by Lemma \ref{l3.2.2} that ${\hh}={\kk}\otimes
{\ll}_\lambda $ and $\tau =1\otimes \tau _\lambda $. If $\dim(\mathfrak
g_{1\lambda })$ is odd, we choose $\kk$ purely even. Then
$1\otimes \tau_\lambda (hX)=\xi (h)[1\otimes \tau_\lambda (X)]
\xi (h)^{-1}$. But the same relation is true if we replace
$\xi$ by $1\otimes \kappa_\lambda $. So
if $\xi_1=[1\otimes \kappa_\lambda ]^{-1}\xi$,
then $\xi _1 (h)$ is even or odd according to the grading
of $\kappa_\lambda (h)$, and commutes with $1\otimes \tau_\lambda $.
Hence $\xi_1$ is of the form $R'\otimes 1$
for a Borel
map  $R'$ from $L^\lambda _0$ into the unitary group of
${\kk}$. Thus
$$
\xi (h)=[1\otimes \kappa_\lambda (h)]
[R'(h)\otimes 1].
$$
The two factors on the right side of this equation commute;
the left side is an even UR and the
first factor on the right is
a unitary $\mu_\lambda$-representation of $L^\lambda _0$,
which is graded or even according to $\dim(\mathfrak g_{1\lambda })$ even or
$\dim(\mathfrak g_{1\lambda })$ odd. So $R'$ is a $\mu_\lambda^{-1}
=\mu_\lambda$-representation of $L^\lambda _0$ in ${\kk}$,
which is graded or even according to $\dim(\mathfrak g_{1\lambda })$ even or
$\dim(\mathfrak g_{1\lambda })$ odd.
This finishes the proof.
\end{proof}
\begin{remark}
 If $\lambda =0$, then $\Phi_\lambda =0,
{\ll}_\lambda =0$, and $\theta _{r0}=(r, 0)$.

\bigskip\noindent
\end{remark}
\begin{remark}
\label{r2}
 For the super Poincar\'e groups
we shall see in the next subsection that the situation
is much simpler
and $L^\lambda_0$ is always connected and satisfies
the conditions of Lemma \ref{l3.2.4}.
\end{remark}

Combining Theorems \ref{t3.1.1} and \ref{t3.2.7} we obtain
the following theorem. Let $\Theta _{r\lambda }$
be the UR of $(G_0, \mathfrak g)$ induced by
$\theta _{r\lambda }$ as described in
Theorem \ref{t3.2.7}.
\begin{theorem}
\label{t3.2.8}
 Let $\lambda $ be such that
$\Phi_\lambda \ge 0$. The assignment that takes $r$ to the {\rm UR}
$\Theta _{r\lambda }$ is an
equivalence of categories
from the category of unitary $\mu_\lambda$-representations
 of $L^\lambda _{00}$ to the
category of {\rm UR}'s of $(G_0, \mathfrak g)$ whose spectra
are contained in the orbit of $\lambda $. In particular,
for $r$ irreducible, $\Theta _{r\lambda }$
is irreducible, and every irreducible {\rm UR} of
$(G_0, \mathfrak g)$ is obtained in this way. If the conditions of
Lemma \ref{l3.2.4} are satisfied, then the $r$'s come from the category
of {\rm UR}'s of $L^\lambda_{0}$.
\end{theorem}

In the case of super Poincar\'e groups (see
Remark \ref{r2} above), $\Theta _{r\lambda}$ induced by
$\theta_{r\lambda}$ represents a
{\it superparticle}. In general the UR
$\pi _{r\lambda }$ of $T_0L_0$
contained in $\Theta _{r\lambda }$ will
{\it not\/} be an irreducible UR of $G_0$.
Its decomposition into irreducibles gives the {\it multiplet\/}
that the UR of $S$
determines. This is of course the set of
irreducible UR's $U_{r\lambda j}$
of $G_0$ induced by the $r_{\lambda j}$
where the
$r_{\lambda j}$ are the irreducible UR's
of $L^\lambda _0$ contained in
$r\otimes \kappa_\lambda$:
$$
r\otimes \kappa_\lambda =\bigoplus r_{\lambda j},
\qquad \pi_{r\lambda }=\bigoplus U_{r\lambda j}.
$$
The set $(r_{\lambda j})$ thus defines the {\it
multiplet\/}. For $r$ trivial the corresponding
multiplet is called {\it fundamental.\/}

\bigskip\noindent
\subsection{The case of the super Poincar\'e groups}
\label{ssec:3.3}
We shall now specialize the entire
theory to the case when $(G_0, \mathfrak g)$ is a {\em super
Poincar\'e group} (SPG). This means that the following
conditions are satisfied.
\begin{itemize}
\item [(a)] $T_0={\real}^{1, D-1}$ is the
$D$-dimensional Minkowski space of signature
$(1, D-1)$ with $D\ge 4$; the Minkowski bilinear
form is $\langle x, x'\rangle =x_0x_0'-
\sum _jx_jx_j'$.

\smallskip\item
[(b)] $L_0={\rm Spin}(1, D-1)$.

\smallskip \item
[(c)] $\mathfrak g_1$ is a real spinorial module for
$L_0$, i.e., is a direct sum of spin representations over
${\complex}$.

\smallskip\item
[(d)] For any $0\not=X\in \mathfrak g_1$, and any
$x\in T_0$ lying in the interior $\Gamma ^+$
of the forward light cone $\Gamma$,
we have
$$
\langle [X,X], x\rangle >0.
$$
\end{itemize}

If in (c) $\mathfrak g_1$ is the sum of $N$ real
irreducible spin modules of $L_0$, we say we are
in the context of {\em $N$-extended supersymmetry}.
Sometimes $N$ refers to the number of
irreducible components over ${\complex}$. In (d)
$$
\Gamma =\{x\ |\ \langle x, x\rangle \ge 0, x_0\ge 0\},
\qquad \Gamma ^+=\{x\ |\ \langle x, x\rangle>0, x_0>0\}.
$$
In the case when $D=4$ and $\mathfrak g_1$ is the Majorana
spinor, the condition (d) is {\it automatic\/}
(one may have
to change the sign of the odd commutators to achieve
this); in the general case, as we shall see below,
it ensures that only positive energy
representations are allowed.

\medskip
We identify $T_0^\ast$ with ${\real}^{1, D-1}$ by the
pairing $\langle x, p\rangle =x_0p_0-\sum _jx_jp_j$.
The dual action of $L_0$ is then the original action.
The orbit structure of $T_0^\ast$ is classical.
\begin{lemma}
\label{l3.3.1}
{\rm (}i{\rm )} Let
$V$ be a
finite dimensional real vector space with a
nondegenerate quadratic form and let $V_1$
be a subspace of $V$ on which the quadratic form
remains nondegenerate. Then the spin representations
of ${\rm Spin}(V)$ restrict on ${\rm Spin}(V_1)$
to direct sums of spin representations of
${\rm Spin}(V_1)$. {\rm (}ii{\rm )} Suppose $V={\real}^{1,D-1}$.
Let $p\in V$
be such that $\langle p, p\rangle =\pm m^2\not=0$
and $V_1=p^\perp$. Then $V_1$ is a quadratic
subspace, the stabilizer $L^p_0$ of $p$ in
${\rm Spin}(V)$ is precisely ${\rm Spin}(V_1)$,
and it is $\simeq {\rm Spin}(D-1)$ for
$\langle p, p\rangle =m^2$ and $\simeq
{\rm Spin}(1, D-2)$ for $\langle p, p\rangle
=-m^2\not=0$.
\end{lemma}
\begin{proof}
(i) Let ${\cc}, {\cc}_1$ be the
Clifford algebras
of $V$ and $V_1$. Then ${\cc}_1^+\subset {\cc}^+$ and
hence, as the spin groups are imbedded in the
even parts of the Clifford algebras, we have
${\rm Spin}(V_1)\subset {\rm Spin}(V)$. Now the spin
modules are precisely the modules for the even parts
of the corresponding Clifford algebras and so, as these
algebras are semisimple, the decomposition of the
spin module of ${\rm Spin}(V)$, viewed as an
irreducible module for ${\cc}^+$, into irreducible modules
for ${\cc}_1^+$ under restriction to ${\cc}_1^+$,
gives the decomposition of the restriction of the original
spin module to ${\rm Spin}(V_1)$. See \cite{D},\cite{V}.

\medskip
(ii) Choose an orthogonal basis $(e_\alpha)_{0\le
\alpha \le D-1}$ such that $\langle e_0, e_0\rangle
=-\langle e_j, e_j\rangle =1$ for $1\le j\le D-1$. It
is easy to see that we can move $p$ to either
$(m,0,\dots ,0)$ or $(0,m,0,\dots ,0)$ by $L_0$ and
so we may assume that $p$ is in one of these two
positions. For $u\in C(V)^+$ it is then a
straightforward matter to verify that $up=pu$
if and only if $u\in C(V_1)^+$. From the
characterization of the spin group (\cite{D},\cite{V})
it is now clear that $L^p_0={\rm Spin}(V_1)$.
\end{proof}
\begin{lemma}
\label{l3.3.2}
 Let $M$ be a connected real
semisimple Lie group whose universal cover does not
have a compact factor, i.e., the Lie algebra of
$M$ does not have a factor Lie algebra whose group
is compact. Then $M$ has no nontrivial morphisms
into any compact Lie group, and hence no nontrivial
finite dimensional {\rm UR}'s.
\end{lemma}
\begin{proof}
We may assume that $M$ is
simply connected. If such a morphism exists we have a
nontrivial morphism $\mathfrak m\longrightarrow \mathfrak k$
where $\mathfrak m$ is the Lie algebra of $M$ and
$\mathfrak k$ is the Lie algebra of a compact Lie group.
Let $\mathfrak a$ be the kernel of this Lie algebra morphism.
Then $\mathfrak a$ is an ideal of $\mathfrak m$ different
from $\mathfrak m$, and so we can write $\mathfrak m$
as $\mathfrak a\times \mathfrak k'$ where $\mathfrak k'$ is also
an ideal and is non zero; moreover, the map
from $\mathfrak k'$ to $\mathfrak k$ is injective. $\mathfrak k'$
is semisimple and admits an invariant negative
definite form (the restriction from the Cartan-
Killing form of $\mathfrak k$), and so its associated
simply connected group $K'$ is
compact. If $A$ is the simply connected group for
$\mathfrak a$, we have $M=A\times K'$, showing that $M$ admits
a compact factor, contrary to hypothesis.
\end{proof}
\begin{corollary}
\label{c3.3.3}
 If $V$ is a quadratic vector
space of signature $(p,q), p,q>0, p+q\geq 3$, then ${\rm Spin}(V)$
does not have any nontrivial map into a compact Lie group.
\end{corollary}
\begin{proof}
 The Lie algebra is semisimple and the simple factors are 
 not compact.
\end{proof}
\begin{lemma}
\label{l3.3.4}
We have $\Gamma =\{p\ |\ \Phi_p\ge 0\}$,
i.e., for
any $p\in {\real}^{1, D-1}$,
$$
\Phi_p\ge 0\Longleftrightarrow p_0\ge 0,
\langle p, p\rangle \ge 0.
$$
Moreover,
$$
p_0>0, \langle p, p
\rangle >0\Longrightarrow \Phi_p>0.
$$
\end{lemma}
\begin{proof}
 For $0\not=X\in \mathfrak g_1$
we have $\langle [X, X], x\rangle >0$
for all $x\in \Gamma ^+$ and hence the
inequality is true with $\ge 0$ replacing $>0$
for $x\in \Gamma$. Hence $2 \Phi_p(X,X)=
\langle [X, X], p\rangle
\ge 0$ if $p\in \Gamma$. So
$$
\Gamma \subset \{p\ |\ \Phi_p\ge 0\}.
$$

We shall show next that
$$
\{p\ |\ \Phi_p\ge 0\}
\subset \{p\ |\ \langle p, p\rangle \ge 0 \}.
$$
Suppose on the contrary
that  $\Phi_
p \ge 0$ but $\langle p, p\rangle <0$.
Since $\Phi_p$ is invariant under
$L^p_0$ which is connected, we have a map
$L^p_0\longrightarrow {\rm SO}
(\mathfrak g_{1p})$. Then
$L^p_0={\rm Spin}(V_1)={\rm Spin }(1, D-2)$ by Lemma \ref{l3.3.1}, and
Corollary \ref{c3.3.3} shows that
$L^p_0$ has no
nontrivial
morphisms into any compact Lie group. Hence $L^p_0$
acts trivially on $\mathfrak g_{1p}$.
Since $L^p_0$ is a semisimple group,
$\mathfrak g_{1p}$ can be lifted to an $L^p_0$-invariant subspace of $\mathfrak g_1$. Hence, if
$\mathfrak g_{1p}\not=0$, the action of
$L^p_0$ on $\mathfrak g_1$ must have non zero
trivial submodules. However, by
Lemma \ref{l3.3.1}, the spin modules of ${\rm Spin}
(V)$ restrict on ${\rm Spin}(V_1)$ to direct
sums of spin modules of the smaller group and
there is no trivial module in this decomposition.
Hence $\mathfrak g_{1p}=0$, i.e., $\Phi_p=0$.
Hence $p$ vanishes on $[\mathfrak g_1,
\mathfrak g_1]$. Now $[\mathfrak g_1, \mathfrak g_1]$ is stable
under $L_0$ and non zero, and so must be the
whole of $\mathfrak t_0$. So $p=0$, a contradiction.

\medskip
To finish the proof we should prove that if
$\Phi_p\ge 0$ then $p_0\ge 0$. Otherwise $p_0<0$
and so $-p\in \Gamma$ and so from what we have already
proved, we have $\Phi_{-p}=-\Phi_p\ge 0$. Hence
$\Phi_p=0$. But then as before $p=0$, a contradiction.

\medskip
Finally, if $p_0>0$ and $\langle p, p\rangle>0$, then
$\Phi_p>0$ by definition of the SPG structure.
This completes the proof.
\end{proof}
\begin{theorem}
\label{t3.3.5}
 Let $S=(G_0, \mathfrak g)$ be a
{\rm SPG}. Then all stabilizers are connected and
$$
T_0^+=\{p\ |\ \Phi_p\ge 0\}=\Gamma.
$$
Moreover, $\kappa_p$ is an even {\rm UR}
of $L^p_0$, and the irreducible {\rm UR}'s of $S$
whose spectra
are in the orbit of $p$ are in natural bijection
with the irreducible {\rm UR}'s of $L^p_0$. The
corresponding multiplet is then the set of
irreducible {\rm UR}'s parametrized by the
irreducibles of $L^p_0$
occurring in the
decomposition of $\alpha \otimes \kappa_p$ as
a {\rm UR} of $L^p_0$.
\end{theorem}
\begin{proof}
 In view of Theorem \ref{t3.2.7} and Lemma \ref{l3.3.4} we have $T_0^+=\Gamma$. For
$p\in\Gamma$, the stabilizers are
all known classically.
If $\langle p,
p\rangle >0$, $L^p_0={\rm Spin}(D-1)$;
if $\langle p,
p\rangle =0$ but $p_0>0$, then
$L^p_0={\real}^{D-2}\times '{\rm
Spin }(D-2)$; and for $p=0$, $L^p_0=L_0$.
So, except when $D=4$ and $p$ is non zero and is in
the zero mass orbit, the stabilizer is connected and simply
connected, thus $\kappa_p$ is an even $UR$ of $L^p_0$ by Lemma \ref{l3.2.4}. But
in the exceptional case, $L^p_0=
{\real}^2\times ' S^1$ where $S^1$ is the circle, and
Lemma \ref{l3.2.4}
is again applicable. This finishes the proof.
\end{proof}
\subsection{Determination of $\kappa_p$ and
the structure of the multiplets. Examples}
\label{ssec:3.4}
We have seen that
the multiplet defined by the super particle
$\Theta _{\alpha p}$ is parametrized by
the set of irreducible UR's of $L^p_0$ that
occur in the decomposition of
$\alpha \otimes \kappa_p$. Clearly
it is desirable to determine $\kappa_p$
as explicitly as possible.
We shall do this in what follows.

\medskip
To determine $\kappa_p$ the following lemma is useful.
$(W,q)$ is a positive definite quadratic
vector space and $\varphi$ is the bilinear form of $q$.
$C(W)$ is the Clifford algebra of $W$ and $H$
is a connected Lie group with a morphism
$H\longrightarrow {\rm SO}(W)$. $\tau$ is an irreducible
SA representation of $C(W)$ and $\kappa$ is a UR
such that $\kappa(t)\tau(u)\kappa (t)^{-1}=\tau (tu)$
for all $u\in W, t\in H$. {\it We write $\approx$ for
equivalence after multiplying by a suitable
character.}

\begin{lemma}
\label{l3.4.1}
 Suppose that $\dim (W)=2m$
is even and $W_{\complex}:={\complex}\otimes _{\real}W$
has an isotropic subspace $E$
of dimension $m$ stable under $H$.
Let $\eta$ be the action of $H$ on
$\Lambda (E)$ extending its action on $E$. Then
$$
\kappa \approx \Lambda (E)\approx
\Lambda (E^\ast )\qquad (E^\ast \hbox
{is the complex conjugate of }E).
$$
\end{lemma}
\begin{proof}
 Clearly $E^\ast$ is also isotropic
and $H$-stable. $E\cap E^\ast =0$ as otherwise
$E\cap E^\ast \cap W$ will be a non zero isotropic
subspace of $W$. So
$W_{\complex}=E\oplus E^\ast$. We write $\tau^\prime$ for the
representation of $C(W)$ in $\Lambda (E)$ where
$$
\tau ^\prime(u)(x)=u\wedge x, \qquad
\tau ^\prime(v)(x)=\partial (v)(x)\qquad (u\in E,
v\in E^\ast, x\in \Lambda (E)).
$$
Here $\partial (v)$ is the odd derivation taking
$x\in E$ to $2\varphi (x, v)$. It is then
routine to show that
$$
\eta (t)\tau ^\prime(u)\eta (t)^{-1}=\tau ^\prime(tu)
\qquad (u\in E\cup E^\ast ).
$$
Now $\tau ^\prime$ is equivalent to $\tau$ and so
we can transfer $\eta$ to an action,
written again as $\eta$, of $H$ in the
space of $\tau$ satisfying the above relation
with respect to $\tau$. It is not necessary
that $\eta$ be unitary. But we can normalize
it to be a UR, namely $\kappa (t)=
|\det (\eta(t))|^{-1/\dim (\tau)}
\eta(t)$.
\end{proof}
\begin{remark}
 It is easy to give an independent
argument that $\Lambda (E)\approx \Lambda (E^\ast)$.
For the unitary group ${\rm U}(E)$ of $E$
let $\Lambda _r$
be the representation on $\Lambda ^r(E)$,
and let $\Lambda $ be their direct sum; then
a simple calculation of the characters on the
diagonal
group shows that ${\Lambda _r}^\ast \simeq
\det ^{-1}
\otimes
\Lambda _{n-r}$. Hence ${\Lambda }^\ast \simeq
\det ^{-1}\otimes \Lambda $, showing that
${\Lambda }^\ast
\approx \Lambda $. It is then immediate that this
result remains true for any group
which acts unitarily on $E$.
\end{remark}
\begin{corollary}
\label{c3.4.2}
 The conditions
of the above lemma are met if $W =
A\oplus B$ where $A, B$ are orthogonal
submodules for $H$ which
are equivalent. Moreover
$$
\kappa \approx \Lambda (E)\simeq \Lambda (E^\ast )
\simeq \Lambda (A)
\simeq \Lambda (B).
$$
\end{corollary}
\begin{proof}
Take ON bases $(a_j), (b_j)$ for $A$
and $B$ respectively so that the map $a_j\mapsto b_j$ is
an isomorphism of $H$-modules. If $E$ is the span of the
$e_j=a_j+ib_j$, it is easy to check that $E$ is isotropic,
and is a module for $H$ which is equivalent to $A$ and $B$.
\end{proof}

We now assume that for some $r\ge 3$ we have a map
$$
H\longrightarrow {\rm Spin}(r)\longrightarrow
{\rm Spin}(W)
$$
where the first map is surjective, and $H$ acts on $W$
through ${\rm Spin}(W)$.
Further let the
representation of ${\rm Spin}(r)$
on $W$ be spinorial. We write
$\sigma _r$ for the (complex) spin representation of
${\rm Spin}(r)$ if $r$ is odd and $\sigma _r^\pm$ for
the (complex) spin representations of ${\rm Spin}(r)$ if
$r$ is even. Likewise we write $s_r, s_r^\pm$ for
the real irreducible spin modules. Note that $\dim (W)$
must be even.

\begin{lemma}
\label{l3.4.3}
 Let the
representation of ${\rm Spin}(r)$ on $W$
be spinorial. Let $n$ be the number of real
irreducible constituents of $W$ as a
module for ${\rm Spin}(r)$, and, when
$r$ is even, let $n^\pm$ be the
number of irreducible constituents of
real or quaternionic type. We then have
the following determination of $\kappa$.
$$
\begin{array}{lcl}
r \hbox { mod } 8 &\quad& \kappa \\
&& \\
0(n^\pm \hbox { even }) &\quad& \Lambda
\bigl (((n^+/2)\sigma_r^+\oplus (n^-/2)
\sigma_r^-)\bigr )\\
1, 7 (n\hbox { even}) &\quad&\Lambda
\bigl ((n/2)\sigma_r\bigr ) \\
2, 6 &\quad& \Lambda \bigl (n\sigma_r^+\bigr )\approx
\Lambda \bigl (n\sigma_r^-\bigr ) \\
3, 5 &\quad& \Lambda \bigl (n\sigma_r\bigr ) \\
4 & &\Lambda \bigl (n^+\sigma_r^+
\oplus n^-\sigma_r^-\bigr )
\end{array}
$$
\end{lemma}
\begin{proof}
 This is a routine application of the
the Lemma and Corollary above if we note the
following facts.

$r\equiv 0$ : Here $\sigma_r^\pm=s_r^\pm$
and $W=n^+s_r^++n^-s_r^-$.

$r\equiv 1, 7$ : Here $\sigma_r=s_r, W=ns_r$.

$r\equiv 2, 6$ : Over ${\complex}$, $s_r$ becomes
$\sigma_r^+\oplus
\sigma_r^-$
while $\sigma_r^\pm$ do not admit
a non zero
invariant form. So $W_{\complex}=E\oplus E^\ast$
where $E=n\sigma_r^+, E^\ast =n\sigma_r^-$,
and $q$ is zero on $E$.

\smallskip
$r\equiv 3,5$ : $s_r$ is quaternionic,
$W=ns_r$, $W_{\complex}=2n \sigma_r$ and $\sigma_r$
does not admit
an invariant symmetric form.

\smallskip
$r\equiv 4$ : $s_r^\pm$ are quaternionic
and
$\sigma_r^\pm$ do not admit a non zero invariant symmetric
form; $W_{\complex} = E\oplus E^{\ast}$, where $E=n^+\sigma_r^++n^-\sigma_r^-$ and $q$ is zero on $E$.

\medskip\noindent
In deriving these the reader should use the results in \cite{D}
and \cite{V} on the reality of the complex spin modules and the theory
of invariant forms for them.
\end{proof}
\subsubsection{Super Poincar\'e group associated to
${\real}^{1, 3}$: N=1
supersymmetry}
 Here $T_0={\real}^{1,3}, L_0={\rm SL
}(2, {\complex})_{\real}$ where the suffix ${\real}$ means that
the complex group is viewed as a real Lie group. Let
$\mathfrak s={\bf 2}\oplus \overline {\bf 2}$, ${\bf 2}$
being the holomorphic representation of $L_0$ in
${\complex}^2$ and $\overline {\bf 2}$ its complex conjugate.
Thus we identify $\mathfrak s$ with ${\complex}^2\oplus {\complex}^2$
and introduce the conjugation on $\mathfrak s$ given by
$\overline {(u,v)}=(\overline v, \overline u)$.
The action $(u,v)\mapsto (gu,\overline gv)$
of $L_0\ (\overline g$ is the complex conjugate of $g$)
commutes with the conjugation and so
defines the real form $\mathfrak s_{\real}$ invariant under
$L_0$ ({\em Majorana spinor}). We take $\mathfrak t_0$ to be
the space of $2\times 2$ Hermitian matrices and the
action of $L_0$ on it as $g, A\mapsto gA{\overline g}^
{\rm T}$. For $(u_i,\overline {u_i})\in
\mathfrak s_{\real} (i=1,2)$ we put
$$
[(u_1,\overline {u_1}),(u_2,\overline {u_2})]
={1\over 2}(u_1\overline {u_2}^{\rm T}+
u_2\overline {u_1}^{\rm T}).
$$
Then $\mathfrak g=\mathfrak g_0\oplus \mathfrak g_1$ with
$\mathfrak g_0=\mathfrak t_0\oplus \mathfrak l_0,\
\mathfrak g_1=\mathfrak s_{\real}$ is a super Lie algebra and
$(T_0L_0, \mathfrak g)$ is the SLG with which we
are concerned.

\medskip
Here ${\real}^{1,3}\simeq \mathfrak t_0$
by the map $a\mapsto h_a=
\begin{pmatrix}
a_0+a_3 & a_1-ia_2\cr
a_1+ia_2 & a_0-a_3\cr
\end{pmatrix}
$; $\mathfrak t_0\simeq
\mathfrak t_0^\ast$ with $p\in \mathfrak t_0$
viewed as the linear form
$a\mapsto \langle a,p\rangle =
a_0p_0-a_1p_1-a_2p_2-a_3
p_3$. Then
$$
Q_p((u, \overline u))=
{1\over 4}{\overline u}^{\rm T}
h_{\check p}u, \qquad \check p=
(p_0, -p_1, -p_2, -p_3).
$$

I : $p_0>0,\  m^2=\langle p, p\rangle >0$.
We take $p=mI$ so that $L^p_0={\rm SU}(2)$.
Take $E=\{(u,0)\}, E^\ast=\{(0, \overline u)\}$.
Then we are in the set up of Lemma \ref{l3.4.1}. Then
$$
\kappa_p=\Lambda (E)\simeq 2D^0\oplus D^{1/2},
\qquad D^j\otimes \Lambda (E)=
\begin{cases}
2D^j\oplus D^{j+1/2}
\oplus D^{j-1/2}\quad (j\ge 1/2)\cr
2D^0\oplus D^{1/2}\quad (j=0).\cr
\end{cases}
$$
Thus the multiplet with mass $m$
has the same mass
$m$ and spins
$$
\begin{cases}
\{j, j , j+1/2, j-1/2\} (j>0)\cr
\{0, 0, 1/2\} (j=0)
\end{cases}
$$

II : $p_0>0,\ \langle p, p\rangle =0$.
Here we take $p=(1, 0, 0, -1), h_p=
\begin{pmatrix}
0&0\cr 0&2\cr
\end{pmatrix}.$ Then
$L^p_0=\begin{pmatrix}
a&0\cr c&\overline a\cr
\end{pmatrix}$. The
characters $\chi_{n/2}: a\mapsto a^n\ (n\in {\z})$
are viewed as characters of $L^p_0$. Here
$Q_p((u, \overline u))={1\over 4}
{\overline u}^{\rm T}h_{\check p}u=|u_1|^2$. The
radical of $\Phi_p$ is the span of $(e_2, e_2)$
and $(ie_2, -ie_2)$, $e_1, e_2$ being the standard
basis of ${\complex}^2$. We identify $\mathfrak g_{1p}$
with the span of $(e_1, e_1)$
and $(ie_1, -ie_1)$. We now apply Lemma \ref{l3.4.1} with
$E={\complex}(e_1, 0)$ which carries the character defined by
$\chi _{1/2}$; then
$$
\Lambda (E)=\chi _0\oplus
\chi_{1/2}, \qquad \chi_{n/2}\otimes \Lambda (E)
=\chi_{n/2}\oplus
\chi_{(n+1)/2}.
$$
The multiplet is $\{n/2, (n+1)/2\}$. These results
go back to \cite{SS}.

\subsubsection{Extended supersymmetry} Here the SLG has still
the Poincar\'e group as its even part but $\mathfrak g_1$
is the sum of $N>1$ copies of $\mathfrak s_{\real}$. It is
known (\cite{D},\cite{V})that one can identify
$\mathfrak g_1$ with the
direct sum $\mathfrak s_{\real}^N$ of $N$ copies of
$\mathfrak s_{\real}$ in such a way that for the odd
commutators we have
$$
[(s_1,s_2,\dots ,s_N), (s_1',s_2',\dots ,s_N')]=
\sum _{1\le i\le N}[s_i,s_i']^1,
$$
so that
$$
Q_\lambda ((s_1,\dots ,s_N))=\sum _{1\le i\le N}
Q^1_\lambda  ((s_i, s_i)).
$$
Here the index $1$ means the $[\ ,\ ]$ and
$Q$ for the case
$N=1$ discussed above. Let $E^N=NE^1$.

I : $p_0>0, m^2=\langle p ,p\rangle >0$.
Then we apply Lemma \ref{l3.4.1} with $E=E^N$ so that
$\kappa_p=\Lambda (ND^{1/2})$. The
decomposition of the exterior algebra of
$ND^{1/2}$ is tedious but there is no
difficulty in principle. We have
$$
\kappa_p=\sum _{0\le r\le N} c_{Nr}D^{r/2}
\quad c_{Nr}>0,\quad c_{NN}=1.
$$
Then $j+N/2$ is the maximum
value of $r$ for which $D^r$ occurs in
$D^j\otimes \Lambda (ND^{1/2})$. The multiplet
defined by the super particle of mass $m$
is thus
$$
\begin{cases}
\{j-N/2, j-N/2+1/2, \dots , j+N/2-1/2, j+N/2\} \ (j\geq N/2)\cr
\{0,1/2, \dots , j+N/2-1/2, j+N/2\}\ (0\leq j<N/2)\cr
\end{cases}
$$

II : $p_0>0, m=0$. Here
$$
\kappa_\lambda =\Lambda (N\chi_{1/2})=
\sum _{0\le r\le N} {N\choose r}\chi_{r/2}.
$$
The multiplet of the super particle
has the helicity
content
$$
\{r/2, (r+1)/2, \dots , (r+N)/2\}.
$$

\subsubsection{Super particles of infinite spin} The little
groups for zero mass have irreducible UR's which are
infinite dimensional. Since $L^p_0$ is also
a semidirect product its irreducible UR's can be
determined by the usual method. The orbits of $S^1$
in ${\complex}$ (which is identified with its dual) are
the circles $\{|a|=r\}$ for $r>0$ and the stabilizers
of the points are all the same, the group $\{\pm 1\}$.
The irreducible UR's of infinite dimension can
then be parametrized as $\{\alpha _{r,\pm}\}$. Now
$\Lambda (E)=\chi_0\oplus \chi_{1/2}$ and an
easy calculation gives
$$
\alpha_{r,\pm }\otimes \Lambda (E)=
\alpha _{r, +}\oplus \alpha_{r, -}.
$$
The particles in the multiplet with mass $0$ corresponding
to spin
$(r, \pm)$ consist of both types
of infinite spin with the same $r$.

\subsubsection{Super Poincar\'e groups of Minkowski super
spacetimes of arbitrary dimension} Let $T_0=
{\real}^{1,D-1}$.
We first determine $\kappa_p$ in the massive case.
Here $L^p_0={\rm Spin}(D-1)$ and the form $\Phi_p$
is strictly positive definite. So
$\mathfrak g_{1p}=\mathfrak g_1$ and $L^p_0$
acts on it by restriction,
hence spinorially by Lemma \ref{l3.3.1}.
So Lemma \ref{l3.4.3} applies at once. It
only remains to determine $n,n^\pm$ in terms of the
corresponding $N, N^\pm$ for $\mathfrak g_1$ viewed as
a module for $L_0$. Notation is as in Lemma \ref{l3.4.3},
and res is restriction to $L^p_0$; $r=D-1$. This
is done by writing
$\mathfrak g_1$ as a sum of the $s_D$ and determining
the restrictions of the $s_D$ to $L^p_0$ by
dimension counting. We again omit the details but
refer the reader to \cite{D},\cite{V}.

\begin{proposition}
\label{p3.4.4}
When $p_0>0,
m^2=\langle p, p\rangle >0$, $\kappa_p$, the
fundamental multiplet of the
super particle of mass $m$, is given
according to the
following table:
$$
\begin{array}{lclcl}
D \hbox { mod } 8 &\quad & {\rm res}\ (s_D) &\quad & \kappa_p \\
&&&&\\
 0 & & 2s_{D-1} & & \Lambda \bigl (N\sigma_{D-1}\bigr )  \\
 1(N=2k) & & s_{D-1}^++s_{D-1}^- & &
 \Lambda \bigl (k\sigma_{D-1}^+\oplus
k\sigma_{D-1}^-\bigr ) \\
 2(N=2k) & & s_{D-1} & & \Lambda
\bigl (k\sigma_{D-1}\bigr )
 \\
 3 & & s_{D-1} & & \Lambda \bigl (N\sigma_{D-1}^\pm\bigr ) \\
 4 & & s_{D-1} & & \Lambda \bigl (N\sigma_{D-1}\bigr ) \\
 5 & & s_{D-1}^++s_{D-1}^- & & \Lambda \bigl (N\sigma_{D-1}^+\oplus
N\sigma_{D-1}^-\bigr ) \\
 6 & & s_{D-1} & & \Lambda \bigl (N\sigma_{D-1}) \\
 7 & & 2s_{D-1} & & \Lambda \bigl ((2N)\sigma_{D-1}^\pm)
\end{array}
$$

\end{proposition}

We now extend these results to the case when $p$
has zero mass. Let $V={\real}^{1, D-1}
(D\ge 4)$ and $p\not=0$
a null vector in $V$. Let $e_j$
be the standard basis vectors for $V$ so that
$(e_0,e_0)=-(e_j,e_j)=1$ for
$j=1,\dots ,D-1$. We may assume
that $p=e_0+e_1$. Let $V_1'$ be the span
of $e_j (2\le j\le D-1)$. The signature of $V_1'$
is $(0, D-2)$. Then we have the flag $0\subset
{\real}p\subset p^\perp\subset V$ left stable by the
stabilizer $L^p_0$ of $p$ in $L_0$. The
quadratic form on $p^\perp$ has ${\real}p$ as its
radical and so induces a nondegenerate form on
$V_1:=p^\perp/{\real}p$. Write $L_1={\rm Spin}(V_1)$.
Note that $V_1'\simeq V_1$.

\medskip
We have a map $x\mapsto x'$
from $L^p_0$ to $L_1$ where, for
$v\in p^\perp$
with image $v'\in V_1$, $x'v'=(xv)'$. It is known
that this is surjective and its kernel $T_1:=T^p_0$
is isomorphic to $V_1$ canonically: for $x\in T^p_0$,
the vector $xe_0-e_0\in p^\perp$,
and the map that sends $x$ to
the image $t(x)$ of $xe_0-e_0$ in $V_1$ is well defined
and is an isomorphism of $T^p_0$ with $V_1$. The map
$x\mapsto (t(x), x')$ is an isomorphism of $L^p_0$
with the semidirect product $V_1\times '
L_1$. The Lie algebra of the
big spin group $L_0$
has the $e_re_s (r<s)$ as basis and it is
a simple
calculation that the Lie algebra of $L^p_0$
has as basis
$t_j=(e_0+e_1)e_j (2\le j\le D-1),
e_re_s (2\le r<s\le D-1)$ with the $t_j$
forming a basis
of the Lie algebra of $T_1$.
The $e_re_s (2\le r<s\le D-1)$
span a Lie subalgebra of the Lie algebra
of $L_0$ and the corresponding
subgroup $H\subset L^p_0$ is such
that $L^p_0\simeq T_1\times 'H$. For all of this see
\cite{V}, pp. 36-37.

\medskip
We shall now determine
the structure of the restriction to $L^p_0$ of the
irreducible spin representation(s) of $L$,
{\it over ${\real}$ as well as over ${\complex}$.\/}
Since this may not be known widely
we give some details. We begin
with some preliminary remarks.

\medskip
Let $U$ be any
finite dimensional
complex $L^p_0$-module. Write
$U=\oplus _\chi U_\chi$ where $U_\chi$, for any character
(not necessarily unitary) $\chi$ of $T_1$,
is the subspace of all elements
$u\in U$ such that $(t-\chi (t))^mu=0$ for
sufficiently large $m$. The action of $L_1$
permutes the $U_\chi$, and so,
since $L_1$ has no finite nontrivial orbit in the
space of characters of $T_1$,
it follows that the spectrum of $T_1$ consists only of the
trivial character, i.e., $T_1$ acts
{\it unipotently.\/}
In particular $U_1$, the subspace of $T_1$-invariant
elements of $U$, is $\not=0$, an assertion which is
then valid for real modules also. It follows that
we have a strictly increasing filtration
$(U_i)_{i\ge 1}$ where $U_{i+1}$ is the preimage in
$U$ of $(U/U_i)_1$. In particular, if $U$ is
semisimple, $U=U_1$.

\begin{lemma}
\label{l3.4.5}
 Let $W$ be an irreducible
real or complex spin module for $L_0$.
Let $W_1$ be the
subspace of all elements of $W$
fixed by $T_1$ and $W^1:=W/W_1$.
We then have the following.

\item {{\rm (}i{\rm )}} $0\not=W_1\not=W$,
$W_1$ is the unique
proper non zero $L^p_0$-submodule of $W$, and
$T_1$ acts trivially also on
$W^1$.

\smallskip\item {{\rm (}ii{\rm )}} $W_1$ and $W^1$
are both irreducible $L^p_0$-modules
on which $L^p_0/T_1\simeq L_1$ acts as
a spin module.

\smallskip\item {{\rm (}iii{\rm )}} The exact sequence
$$
0\longrightarrow W_1\longrightarrow W
\longrightarrow W/W_1\longrightarrow 0
$$
does not split.

\smallskip\item {{\rm (}iv{\rm )}}
Over ${\real},\ W,W_1,W^1$ are
all of the same type. If
$\dim (V)$ is odd, $W_1\simeq W^1$. Let $\dim (V)$
be even; then, over ${\complex}$,
$W_1, W^1$ are the two irreducible spin
modules for $L_1$; over ${\real},\ W_1\simeq W^1$
when $W$ is of complex type, namely, when
$D\equiv 0, 4\ {\rm mod}\ 8$; otherwise, the
modules $W_1, W^1$ are the two irreducible
modules of $L_1$ {\rm (}which are either real
or quaternionic).
\end{lemma}
\begin{proof}
 We first work over ${\complex}$.
Let $C$ be the Clifford algebra of $V$.
The key point is that $W_1\not=W$.
Suppose $W_1=W$. Then
$t_j=0$ on $W$ for all $j$. If $D$ is odd,
$C^+$ is a full matrix algebra and so all
of its modules are faithful, giving a
contradiction. Let $D$ be even and $W$
one of the spin modules for $L_0$.
We know that
inner automorphism by the invertible
odd element $e_2$ changes $W$ to the other
spin module. But as $e_2t_je_2^{-1}=
t_j$ for $j>2$ and $-t_2$ for $j=2$, it
follows that $t_j=0$ on the other spin
module also. Hence $t_j=0$ in the
irreducible module for the full
Clifford algebra $C$. Now $C$ is isomorphic
to a full matrix algebra and so its modules
are faithful, giving again $t_j\not=0$,
a contradiction.

\medskip
Let $(W_i)$ be the strictly increasing
flag of $L^p_0$-modules, with $T_1$
acting trivially on each $W_{i+1}/W_i$, defined
by the previous discussion. Let
$m$ be such that $W_m=W$. Clearly $m\ge 2$.
On the other hand, the element $-1$ of $L_0$
lies in $L^p_0$ and as it acts
as $-1$ on $W$, it acts as $-1$ on all
the $W_{i+1}/W_i$. Hence $\dim (W_{i+1}/W_i)\ge
\dim (\sigma _{D-2})$ (see Lemma 6.8.1 of \cite{V}),
and there is equality
if and only if $W_{i+1}/W_i\simeq \sigma _{D-2}$.
Since $\dim (\sigma_{D})=
2\dim (\sigma _{D-2})$, we see at once that
$m=2$ and that both
$W_1$ and $W/W_1$ are irreducible
$L^p_0$-modules which are
spin modules for $L_1$. The exact sequence
in (iii) cannot split, as otherwise $T_1$
will be trivial on all of $W$.
Suppose now $U$ is a non zero
proper $L^p_0$-submodule of $W$. Then $\dim (U)=
\dim (W_1)$ for the same dimensional argument as above, and so $U$ is irreducible, thus
$T_1$ is trivial on it, showing that
$U=W_1$. We have thus proved (i)-(iii).

\medskip
We now prove (iv). There is nothing to prove
when $D$ is odd since there is only one spin
module. Suppose $D$ is even. Let us again write
$\sigma_{D-2}^\pm$ for the $L^p_0$-modules
obtained by lifting the irreducible
spin modules of $L_1$ to $L^p_0$. We consider
two cases.

\medskip\noindent
${\it D\equiv 0, 4\ \hbox {\rm mod}\ 8}$ : In this
case $\sigma _D^\pm$ are self dual while
$\sigma _{D-2}^\pm$ are dual to each other. It is not restrictive to assume $W=\sigma _D^+$ and $W_1=\sigma _{D-2}^+$.
We have the quotient map $W=\sigma _D^+\longrightarrow W^1=\sigma$; and $\sigma$
is to be determined. Writing $\sigma ^\prime$
for the dual of $\sigma$, we get
$\sigma ^\prime\subset (\sigma _D^+)^\prime
\simeq\sigma _D^+$, so that, by
the uniqueness of the submodule proved above,
$\sigma ^\prime =\sigma _{D-2}^+$. Hence
$\sigma =\sigma _{D-2}^-$.

\medskip\noindent
${\it D\equiv 2, 6\
 \hbox {\rm mod}\ 8}$ : Now
$\sigma _D^\pm$ are dual to each other while
$\sigma _{D-2}^\pm$ are self dual. Again, suppose $W=\sigma _D^+$ and $W_1=\sigma _{D-2}^+$. The above
argument then gives $\sigma ^\prime
\subset \sigma _D^-$.
On the other hand, the inner automorphism by
$e_2$ transforms $\sigma _D^+$
into $\sigma _D^-$,
and the subspace of $\sigma _D^+$ fixed by
$T_1$ into the corresponding subspace of
$\sigma _D^-$, while
at the same time changing $\sigma _{D-2}^+$
into $\sigma _{D-2}^-$. Hence it changes
the inclusion
$\sigma _{D-2}^+\subset \sigma _D^+$
into the inclusion
$\sigma _{D-2}^-\subset \sigma _D^-$. Hence
we have $\sigma ^\prime =\sigma _{D-2}^-$. Dualizing,
this gives $\sigma =\sigma _{D-2}^-$ once again.
This finishes the proof of the lemma over ${\complex}$.

\medskip
We now work over ${\real}$. Since both $V$ and $V_1$
have the same signature $D-2$, it is immediate
that the real
spin modules for
$L_0$ and $L_1$ are of the same type.
As an $L_0$-module, the
complexification $W_{\complex}$
of $W$ is either irreducible or is a direct
sum $U\oplus \overline U$ where $U$ is a complex
spin module for $L_0$. In the first case
$W$ is of real type and the lemma follows from
the lemma for the complex spin modules. In the
second case $W$ is of quaternionic or complex
type according as $U$ and $\overline U$ are
equivalent or not.

\medskip
{\it Complex type\/} : We have $W_{\complex}
=U\oplus Z$ where $Z=\overline U$
is the complex
conjugate of $U$. We have $U\simeq
\sigma _D^+, Z\simeq \sigma _D^-$.
Since ${\complex}W_1=U_1\oplus Z_1$ it is
clear that $0\not=W_1\not=W$. The real
irreducible spin modules of $L_1$
have dimension $2^{D/2-1}$ and so
we find that $\dim (W_1)=
2^{D/2-1}$ and $W_1, W^1$ are
both irreducible; they are equivalent
as they are of complex type. A non
zero proper submodule $R$ of $W$ then has
dimension $2^{D/2-1}$ and so must
be irreducible. Hence either
$R=W_1$ or $W=W_1\oplus R$. But then
$W=W_1$, a contradiction. The same argument
shows that the exact sequence in the
lemma does not split.

\medskip
{\it Quaternionic type\/} : For proving
(i)-(iii) of the lemma the argument
is the same as in the complex type,
except that $Z\simeq U$. We now check (iv).
The case of odd dimension is obvious. So
let $D$ be even and $W_1\simeq s_{D-2}^+$.
Then ${\complex}W_1\simeq 2\sigma _{D-2}^+$
so that $U_1\simeq Z_1\simeq
\sigma _{D-2}^+$. Then $U^1\simeq Z^1\simeq
\sigma _{D-2}^-$ and so ${\complex}W^1\simeq
2\sigma _{D-2}^-$. Hence $W^1\simeq
s_{D-2}^-$.

\medskip
The lemma is completely proved.
\end{proof}
\begin{remark}
 Since we are interested in the
quotient $W^1$ rather than $W_1$ below, we change
our convention slightly; for $W=s_D^\pm$ we
write $W^1=s_{D-2}^\pm$ and $W_1=s_{D-2}^\mp$.
\end{remark}
We now come to the discussion of the
structure of $\kappa_p$ when $p$ is
in a massless orbit.
\begin{proposition}
\label{p3.4.6}
 Let $p_0>0, \langle
p, p\rangle =0$. Then ${\rm rad}\ (\Phi_p)=
\mathfrak g_1^{T_1}$, the subspace of elements
of $\mathfrak g_1$ fixed by $T_1$. Moreover
$T_1$ acts trivially on $\mathfrak g_{1p}$,
${\rm rad}\ (\Phi_p)\simeq \mathfrak g_{1p}$
except in the cases $D\equiv 2, 6\ {\rm mod}\ 8$
when ${\rm rad}\ (\Phi_p)$ and $\mathfrak g_{1p}$
are dual to each other, and
$L^p_0/T_1\simeq L_1$ acts spinorially on
${\rm rad}\ \Phi_p$ and $\mathfrak g_{1p}$.
In all cases $\dim ({\rm rad}\ (\Phi_p))=
\dim (\mathfrak g_{1p})=(1/2)\dim (\mathfrak g_1)$.
For $\mathfrak g_{1p}$ as well as the associated
$\kappa_p$ the results are as in the following
table.
$$
\begin{array}{lclcl}
D \hbox { mod } 8&\quad &\mathfrak g_{1p}&\quad & \kappa_p \\
&&&& \\
 0, 4 &\quad & Ns_{D-2} &\quad & \Lambda \bigl
(N\sigma_{D-2}^\pm\bigr ) \\
 2(N^\pm =2n^\pm) &\quad & N^+s_{D-2}^+
+N^-s_{D-2}^- &\quad &
 \Lambda \bigl (n^+\sigma _{D-2}^++
n^-\sigma_{D-2}^-\bigr )
 \\
 6 &\quad & N^+s_{D-2}^++N^-s_{D-2}^- &\quad & \Lambda
\bigl (N^+\sigma _{D-2}^++N^-\sigma _{D-2}^-\bigr ) \\
 1, 3 (N=2n) &\quad & Ns_{D-2} &\quad & \Lambda \bigl
(n\sigma _{D-2}\bigr ) \\
 5, 7 &\quad & Ns_{D-2} &\quad & \Lambda \bigl (N\sigma _{D-2}
\bigr )
\end{array}
$$
\end{proposition}
\begin{proof}
We have $\mathfrak g_1=\oplus _
{1\le i\le N}\mathfrak h_i$ where the $\mathfrak h_i$
are real irreducible spin modules and
$[\mathfrak h_i, \mathfrak h_j]=0$ for $i\not=j$ while
$\langle [X, X], q\rangle >0$ for all
$q\in \Gamma ^+$. Let $\mathfrak r_p=
{\rm rad}\ \Phi_p$ and $\mathfrak r_{ip}$
the radical of the restriction of $\Phi_p$
to $\mathfrak h_i$. Since $Q_p(X)=
\sum _i Q_p(X_i)$ where $X_i$ is the
component of $X$ in $\mathfrak h_i$, it
follows that $\mathfrak r_p=\oplus _i\mathfrak r_{ip}$.
We now claim that $\mathfrak r_{ip}=(\mathfrak h_i)_1$,
namely, the subspace of elements of
$\mathfrak h_i$ fixed by $T_1$. Since $\mathfrak r_{ip}$
is a $L^p_0$-submodule it suffices, in view of
the lemma above, to show that $0\not=
\mathfrak r_{ip}\not=\mathfrak h_i$. If $\mathfrak r_{ip}$
were $0$, $\Phi_p$ would be strictly positive
definite on $\mathfrak h_i$, and hence the action
of $L^p_0$ will have an invariant positive
definite quadratic form. So the action of
$L^p_0$ on $\mathfrak h_i$ will be semisimple,
implying that $T_1$ will act trivially on
$\mathfrak h_i$. This is impossible, since,
by the preceding
lemma, $\mathfrak h_i\not=(\mathfrak h_i)_1$.
If $\mathfrak r_{ip}=\mathfrak h_i$, then
$\Phi_p=0$ on $\mathfrak h_i$, and this will
imply that $p=0$. Thus $\mathfrak r_{ip}=
(\mathfrak h_i)_1$, hence $\mathfrak r_p=
(\mathfrak g_1)_1$. The other assertions except
the table are now clear. For the table we
need to observe that $\mathfrak g_{1p}=
\oplus _i\mathfrak h_i/\mathfrak r_{ip}$ and that
$\mathfrak r_{ip}\simeq \mathfrak h_i/\mathfrak r_{ip}$
except when $D\equiv 2, 6\ {\rm mod}\ 8$;
in these cases, the two modules are the
two real or quaternionic spin modules
which are dual to each other. The table is worked
out in a similar manner to Proposition
\ref{p3.4.4}. We omit the details.
\end{proof}
\begin{remark}
 The result that the dimension of
$\mathfrak g_{1p}$ has $1/2$ the dimension of
$\mathfrak g_1$ extends the known
calculations when $D=4$ (see \cite{FSZ}).
\end{remark}

\subsubsection{The role of the $R$-group in classifying
the states of $\kappa_p$} In the case of $N$-extended
supersymmetry we have two groups
acting on $\mathfrak g_1$: $L^p_0$, the even
part of the little super group at $p$, and
the $R$-group (\cite{D},\cite{V}) $R$. Their actions
commute and they both leave the quadratic form
$Q_p$ invariant. In the massive case we have a map
$$
L^p_0\times R\longrightarrow {\rm Spin}(\mathfrak g_{1p})
$$
so that one can speak of the restriction
$\kappa^\prime_p$ of the spin
representation of ${\rm Spin}(\mathfrak g_{1p})$ to
$L^p_0\times R$. The same is true in the
massless case except we have to replace $L^p_0$
by a two-fold cover of it. It is thus desirable to not
just determine $\kappa_p$ as we have done but
actually determine this representation
$\kappa^\prime_p$ of
$L^p_0\times R$. We have not done this but there is no
difficulty in principle. However, when $D=4$,
we have a beautiful
formula \cite{FSZ}. To describe this, assume that
we are in the massive case. We first remark
that $\mathfrak g_1\simeq \mathbb{H}^N\otimes_{\mathbb{H}}S_0$
where $S_0$ is the quaternionic irreducible
of ${\rm SU}(2)$ of dimension $4$. Thus the $R$-group
is the unitary group ${\rm U}(N, \mathbb{H})$. Over ${\complex}$
we thus have $\mathfrak g_{1{\complex}}\simeq {\complex}^{2N}
\otimes {\complex}^2$ where the $R$-group is the symplectic
group ${\rm Sp}(2N, {\complex})$ acting on the first
factor and $L^p_0\simeq {\rm SU}(2)$ acts as
$D^{1/2}$ on the second factor. The irreducible
representations of ${\rm Sp}(2N, {\complex})
\times {\rm SU}(2)$ are outer tensor products of
irreducibles $a$ of the first factor and $b$ of
the second factor, written as $(a,b)$. Let
${\bf k}$ denote the irreducible of dimension $k$
of ${\rm SU}(2)$ and $[2N]_k$ denote the irreducible
representation of the symplectic group in the space
of traceless antisymmetric tensors of rank $k$ over
${\complex}^{2N}$; by convention for $k=0$ this is the
trivial representation and for $k=1$ it is the vector
representation. Then
$$
\kappa^\prime_p\simeq
([2N]_0,{\bf N+1})+([2N]_1,{\bf N})+\dots ([2N]_k,{\bf
N+1-k})+\dots ([2N]_N,{\bf 1}).
$$
To see how this follows from our theory note that
${\complex}^{2N}\otimes e_1$ is a subspace satisfying the
conditions of Lemma \ref{l3.4.1} for the
symplectic group and so $\kappa_p\simeq
\Lambda ({\complex}^{2N})$. It is known that
$$
\Lambda ({\complex}^{2N})\simeq \sum _
{0\le k\le N}(N+1-k)[2N]_k.
$$
On the other hand we know that the
representations of ${\rm SU}(2)$ in
$\kappa_p$ are precisely the ${\bf N+1-k}
(0\le k\le N)$.
The formula for $\kappa^\prime _p$ is now
immediate. In the massless case the $R$-group
becomes ${\rm U}(N)$ and
$$
\kappa^\prime_p\simeq \sum _{0\le k\le N}
((N-k)/2,[N]_k)
$$
where $r/2$ denotes the character denoted
earlier by $\chi_r$ and $[N]_k$ is the
irreducible representation of ${\rm U}(N)$
defined on the space $\Lambda ^k({\complex}^N)$. We
omit the proof which is similar.

\bigskip\noindent

\begin{acknowledgement}
G.~C. gratefully acknowledges a grant of the Universit\`a di Genova
that has made possible a visit to Los Angeles during which some
of the work for this paper has been done. He thanks Professor V.~S.~Varadarajan for his warm hospitality.\\
V.~S.~V.~would like to thank professor
Giuseppe Marmo and INFN, Naples, professor Sergio Ferrara
and CERN, Geneva, and Professors Enrico Beltrametti and
Gianni Cassinelli and INFN, Genoa, for their hospitality
during the summers of 2003 and 2004, during which most of the
work for this paper was done. We are grateful to Professor Pierre Deligne
of the Institute for Advanced Study, Princeton, NJ, for his interest in our work
and for his comments which have improved the paper.
\end{acknowledgement}


\end{document}